\documentclass[aps,pre,twocolumn]{revtex4}

\usepackage{amsmath}
\usepackage{color}
\usepackage{graphicx}
\usepackage{indentfirst}

\newcommand{\sFrac}[2]{{\textstyle\frac{#1}{#2}}}
\newcommand{\pair}{{(\hskip-0.03cm\alpha\hskip-0.03cm)}}
\newcommand{\generalPair}{{ij}}

\begin{document}
\title{Low-energy non-linear excitations in sphere packings}
\author{Edan Lerner, Gustavo D{\"u}ring, and Matthieu Wyart}
\affiliation{Center for Soft Matter Research, Department of Physics, New York University, New York, NY 10003}
\date{\today}
\begin{abstract}
We study theoretically and numerically how hard frictionless particles in random packings can rearrange. We demonstrate the existence of two distinct unstable non-linear modes of rearrangement, both associated with the opening and the closing of contacts.
Mode one, whose density is characterized by some exponent $\theta'$, corresponds to motions of particles extending throughout the entire system. Mode two, whose density is characterized by an exponent $\theta \ne \theta'$, corresponds to the local buckling of a few particles.
Mode one is shown to yield at a much higher rate than mode two when a stress is applied. We show that the distribution of contact forces follows $P(f)\sim f^{\min( \theta',\theta)}$, and that imposing that the packing cannot be densified further leads to the bounds $\gamma\geq\frac{1}{2+\theta'}$ and $\gamma\geq\frac{1-\theta}{2}$, where $\gamma$ characterizes the singularity of the pair distribution function $g(r)$ at contact. These results extend the theoretical analysis of \cite{theory} where the existence of mode two was not considered. We perform numerics that support that these bounds are saturated with $\gamma\approx0.38$, $\theta\approx0.17$ and $\theta'\approx0.44$. We measure systematically  the stability of all such modes in packings, and confirm their marginal stability. The principle of marginal stability thus allows to make clearcut predictions on the ensemble of configurations visited in these out-of-equilibrium systems, and on the contact forces and pair distribution functions. It also reveals the excitations that need to be included in a description of plasticity or flow near jamming, and suggests a new path to study two-level systems and soft spots in simple amorphous solids of repulsive particles.

\end{abstract}
\maketitle

\section{Introduction}
The dynamics in amorphous materials is so slow that thermal equilibrium cannot be reached. In these systems properties are history-dependent, and configurations of equal energy are not equiprobable. Understanding the properties of these materials is thus a challenge, as it requires  some knowledge on the dynamics and the configuration space it explores. It appears that one key aspect of amorphous solids  is  the possibility for a group of particles to rearrange locally. This phenomenon is believed to govern the low-temperature properties of glasses, where the excitations governing the specific heat are not phonons but two-level systems where a group of particles can switch between two distinct configurations \cite{phillips_book}.   Local rearrangements, the so-called shear-transformation zones,  also dominate the rheological properties of amorphous solids under shear \cite{langer98,argon,04ML}. 
Near the glass transition where structural relaxation is thermally driven, is it also observed that relaxation occurs in favored locations \cite{harrowell2,brito07b}. These soft spots  correspond to regions where low-frequency vibrational modes are quasi-localized 
\cite{manning2}.
 In all these situations elucidating the nature of these local rearrangements, and determining what controls their density, has remained a challenge.

Here we study these questions in packings of hard frictionless spheres, perhaps the simplest model of amorphous solids. It is widely used as a model system for structural glasses and for granular systems, and its microscopic structure has been studied carefully \cite{O'Hern03,donev2,Silbert06,12CCPZ,edan2} . It is found in particular that in such packings there are many particles that are almost touching, but not quite: the distribution function of the gaps between particles $g(h)$ has a singularity near contact $g(h)\sim h^{-\gamma}$, with $\gamma\approx 0.4$ \cite{donev2,Silbert06,12CCPZ}. It has been recently noticed that the  distribution of the contact force amplitude $P(f)$ is also characterized by a non trivial exponent $P(f)\sim f^\theta$ \cite{edan2,12CCPZ} where $\theta\in [0.18,0.45]$, depending on the system preparation. 
Simple toys model \cite{Liu2,sno} that have been proposed early on to compute $P(f)$ give $\theta=0$. In general, computing the structure is complicated because it requires some knowledge on the ensemble of configurations visited by the dynamics. Propositions considering that all mechanically stable states are equally likely~\cite{edwards}, or that packings are maximally random \cite{torquato}
have so far not been able to rationalize these findings. Mean-field approaches based on the replica method \cite{Zamponi,12CCPZ} make quite accurate estimations on the range of values at which jammed packings can be found, but currently predict $\theta=\gamma=0$.

\begin{figure}[ht]
\includegraphics[scale = 0.33]{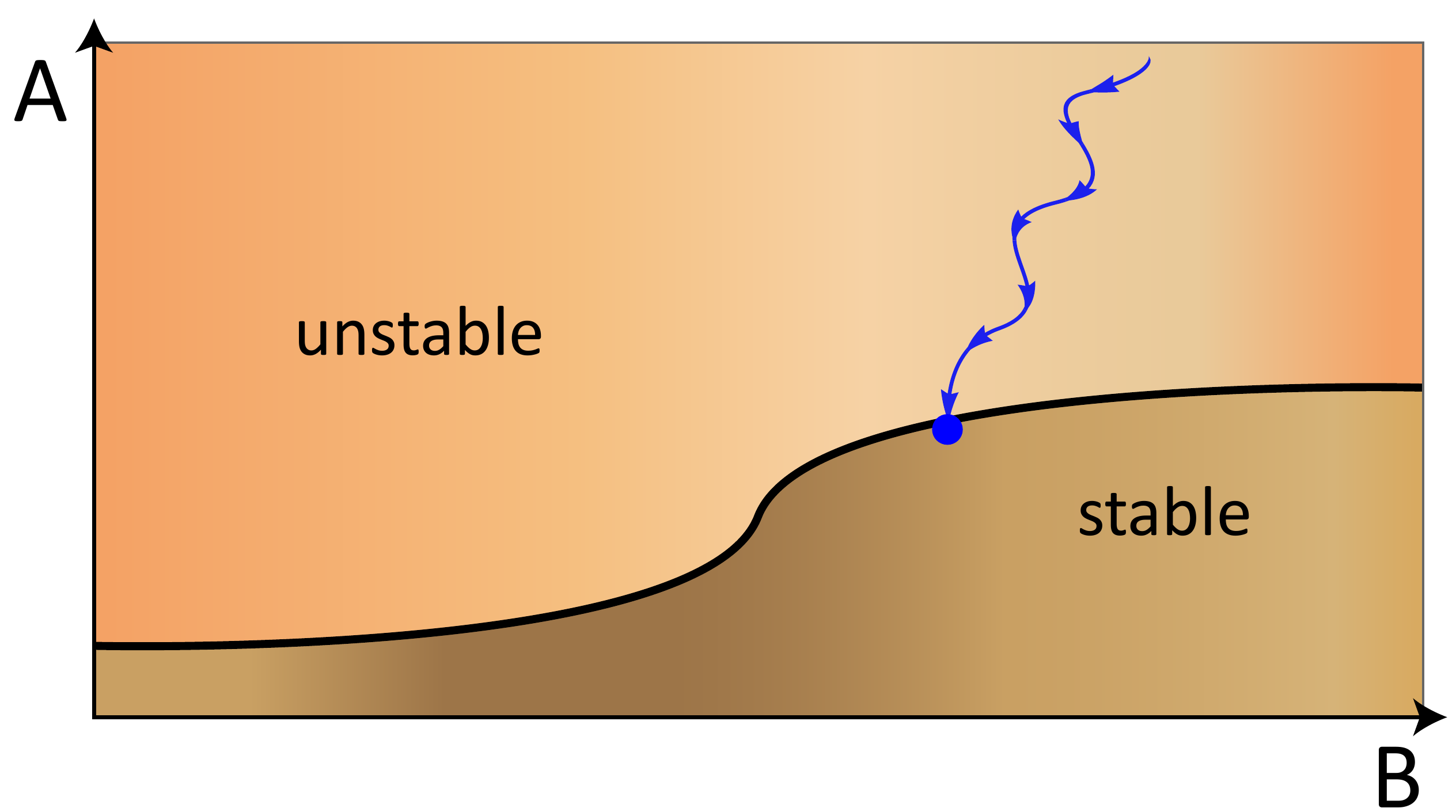}
\caption{Illustration of a marginal stability diagram. $A$ and $B$ are observables characterizing the structure. The thick black
line corresponds to marginal stability: it separates a region where dynamical modes or excitations are stable, 
from a region where these modes are unstable. The arrow-decorated blue line illustrates a dynamical trajectory occurring when the system is prepared. The system is initially unstable, until it reaches the marginality line. At this point if the temperature or the tapping that drives the system is sufficiently small, the dynamics will stop, and the system will lie close to the marginality line. In the case of soft spheres, modes are vibrational, $A$ corresponds to the coordination $z$ and $B$ to the compressive strain $e\sim \phi-\phi_c$ \cite{Wyart052,Wyart053}, where $\phi$ is the packing fraction. In the present work modes are non-linear excitations associated with opening and closing contacts, $A$ corresponds to the exponent $\theta$ or $\theta'$ characterizing some aspects of the force distribution, and $B$ is the exponent $\gamma$ describing the distribution of gaps between particles that are not in contact. On the stability line many soft excitations are present, and rich dynamics, such as crackling noise, can occur.}
\label{marginal}
\end{figure}

Here we shall contend that both the structural properties of packings and their low-energy excitations can be understood in a common framework, based on the principle of marginal stability. This principle, illustrated in Fig.~\ref{marginal}, states that the jammed configurations at which the system arrives are just stable toward the relaxation processes that drive the dynamics in the unjammed phase. If realized, this principle is powerful, both because it  reduces the configurational space of the jammed states encountered, and because it implies an abundance of low-energy excitations that are expected to strongly affect the response of the system. The principle of marginal stability has been successfully applied in some glassy systems with long-range interactions, most importantly in Coulomb glasses where it implies that the density of states  must vanish at the Fermi energy  \cite{efros,monroe,Pazmandi,goethe,markus}, but also  in fully-connected spin glasses \cite{thouless,moore,horner} where it states that the distribution of local fields must vanish at vanishing local fields \cite{thouless}. In both cases, marginality leads to the presence of power-law avalanches of rearrangements under forcing, referred to as crackling noise~\cite{crack}.

Marginal stability has been applied previously to compressed packings of soft particles \cite{Wyart052,Wyart053} and to colloidal glasses \cite{brito07a,brito3}. In these cases linear stability is considered, and the modes at play are simply  vibrational. Requiring that packings lie close to an elastic instability yields $z-z_c\sim   |\phi-\phi_c|^{1/2}$ where $z$ is the coordination, $z_c$ is twice the spatial dimension, and $|\phi-\phi_c|$ is the distance from the jamming threshold, expressed in terms of packing fraction. This scaling is indeed observed numerically \cite{O'Hern03,brito07a} and in emulsions \cite{brujic2}.
It can be shown using variational arguments  \cite{Wyart05,Wyart053} or effective medium approximation  \cite{wyart2010,13DLW} that this behavior of the coordination implies various elastic anomalies and two associated diverging length scales near the jamming threshold, which are indeed observed \cite{revue,vitelli2010,Silbert05,respprl,13DLW,berthier13}. However, this approach is mostly limited to linear properties and cannot explain some aspects of the structure, e.g. the distribution of forces, and leads to a limited insight on the non-linear mechanisms that govern plastic flow  \cite{rouxDL,Schreck}. 

To bridge this gap, one of us  has considered  relaxation mechanisms associated with changes of  contact networks in packings \cite{theory}.
The excitations considered were extended in space, and imposing their stability led to an inequality between the exponent characterizing the distribution of contact forces and the exponent characterizing the distribution of gaps between particles not in contact.
It was assumed that the density of these excitations is controlled by the distribution of contact forces at low-force. Here we revisit this approach and incorporate significant modifications. 
Most importantly, we find that they are two distinct unstable excitations associated with the opening and closing of contacts, instead of one. The novel excitations correspond to local rearrangements of a few particles. 
This finding is important, both because local excitations are expected to affect various properties of amorphous solids as discussed above, but also because, at least for the system preparation we use, these local excitations appear to be more numerous that the extended excitations introduced in  \cite{theory}, and therefore govern the distribution of contact forces. 

This paper is organized as follows. In Section~\ref{smallForces} the relationship between weak contact forces and packing geometry is investigated, and in Section~\ref{stabilityCriterion} the nonlinear stability criterion of  such contacts toward opening is derived. 
In Section~\ref{marginalStability} a scaling analysis of this criterion is performed, which reveals the presence of two relaxation mechanisms, and yields bounds on their respective densities. In Section~\ref{testingRelation} we test numerically our theoretical predictions and establish the existence of two distinct modes of relaxation. We also show that jammed packings produced under our protocol are indeed marginally stable. We demonstrate how our findings imply relations between $P(f)$ and $g(h)$. In Section~\ref{discussion} we argue that marginal stability leads to a simple explanation for the power-law avalanches of plasticity observed in packings under loading \cite{rouxDL}, and in Section~\ref{softParticles} we discuss how our arguments extend to soft compressed particles. Section~\ref{conclusion} summarizes our results and indicates relevant open questions.

\section{Geometric nature of small forces in packings}
\label{smallForces}

Contacts carrying small forces play a special role in packings of hard particles, as they are the most likely to open and lead to rearrangements when external stresses are applied. 
The density of weak forces thus strongly affects how the solid can flow, and it is important to understand what causes certain contacts to carry weak forces. Contact forces can be expressed geometrically in random packings of frictionless particles, as we now recall. 
We shall initially assume  that the packing is contained in a cubic box of  volume $\Omega$ made of rigid walls, and present the main theoretical results in this  context.
Our results are straightforward to extend to periodic boundaries, that we shall use to perform numerical tests.  
We consider a disordered packing of $N$ hard frictionless particles of diameter $a_0$, in spatial dimension $d$. The packing is formed  by pushing particles together by reducing the box size, so as to apply a pressure $p$ that fixes the scale of contact forces in the system.  Microscopically, the boundaries apply external forces ${\vec F}_i$ on all the particles $i$ in contact with it.  
We denote the $i^{\mbox{\tiny th}}$ particle's position by 
$\vec{R}_i$, the pairwise vector of differences by $\vec{R}_\generalPair\equiv\vec{R}_j - \vec{R}_i$, and 
the bare pairwise distance by $r_\generalPair \equiv \sqrt{\vec{R}_\generalPair\cdot\vec{R}_\generalPair}$. 

Mechanical stability in such a packing requires that there exists no floppy modes, i.e.~no collective motions of the degrees of freedom of the system (that include the $Nd$ degrees of freedom of the particles and changes in the box size) for which the distances between objects in contact (including both particles and the box) are fixed. If such a floppy mode existed, the system would flow along it. 
Rapidly compressed, or poly-disperse packings of hard frictionless spherical particles are in fact {\it isostatic} \cite{alexander,revue,tkachenko2,moukarzel,O'Hern03,Wyart053}:  the average number of contacts between particles, known as the coordination number $z$, is \emph{just sufficient} to guarantee mechanical stability and to avoid the presence of floppy modes, corresponding to $z=z_c=2d$  \cite{alexander,tkachenko2,moukarzel}.
\begin{figure}[ht]
\includegraphics[scale = 0.45]{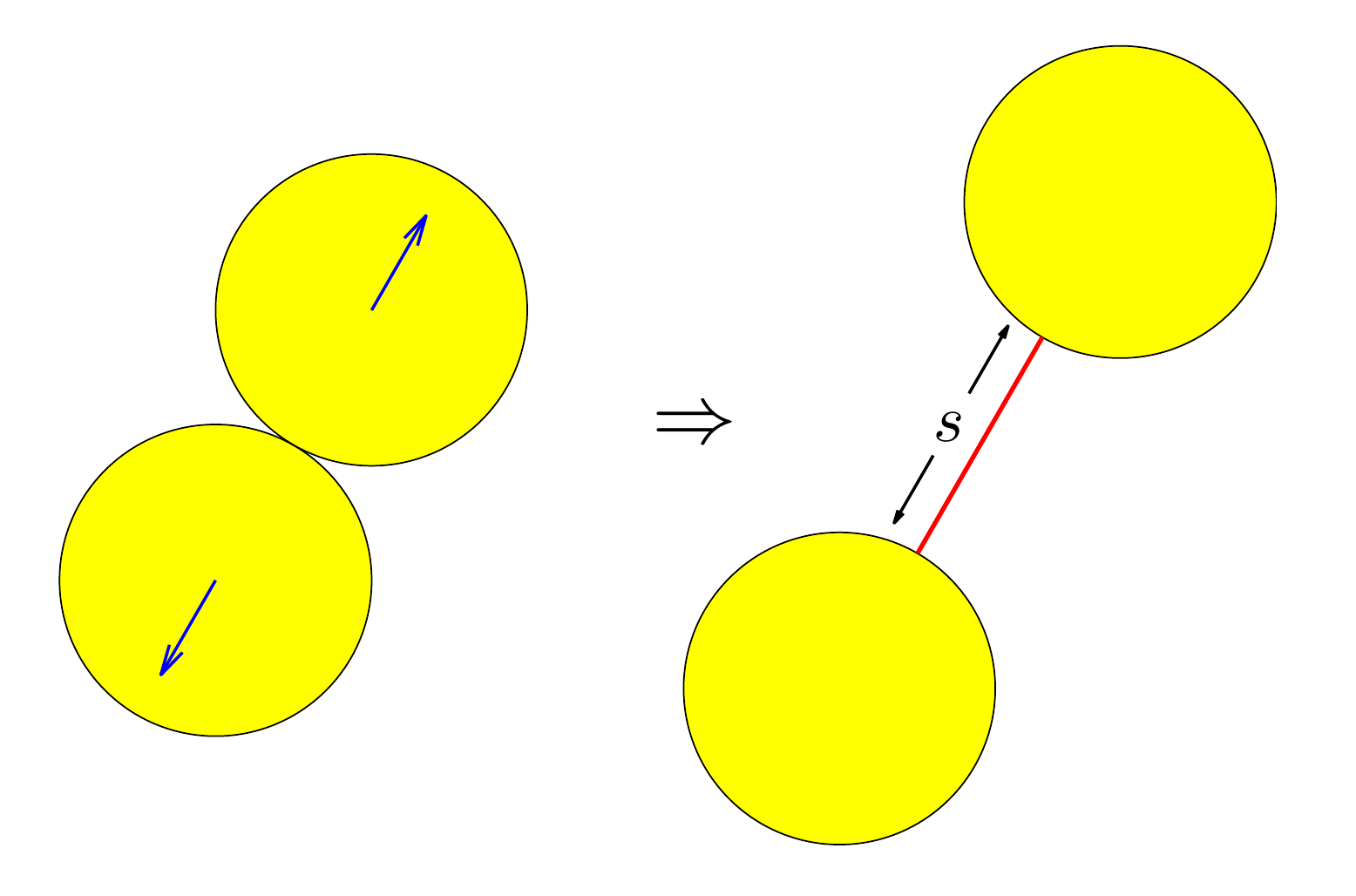}
\caption{Illustration of the perturbation considered in this work. We select the contact $\alpha$ (left panel), and displace
the pair of particles that form the contact $\alpha$ by a distance $s$.}
\label{pushedFig}
\end{figure}
In an isostatic system the removal of any contact leads to the creation of one floppy mode, whose properties govern the magnitude of the force found in the same contact before its removal, as we shall exemplify below.

\begin{figure*}[ht]
\includegraphics[scale = 0.48]{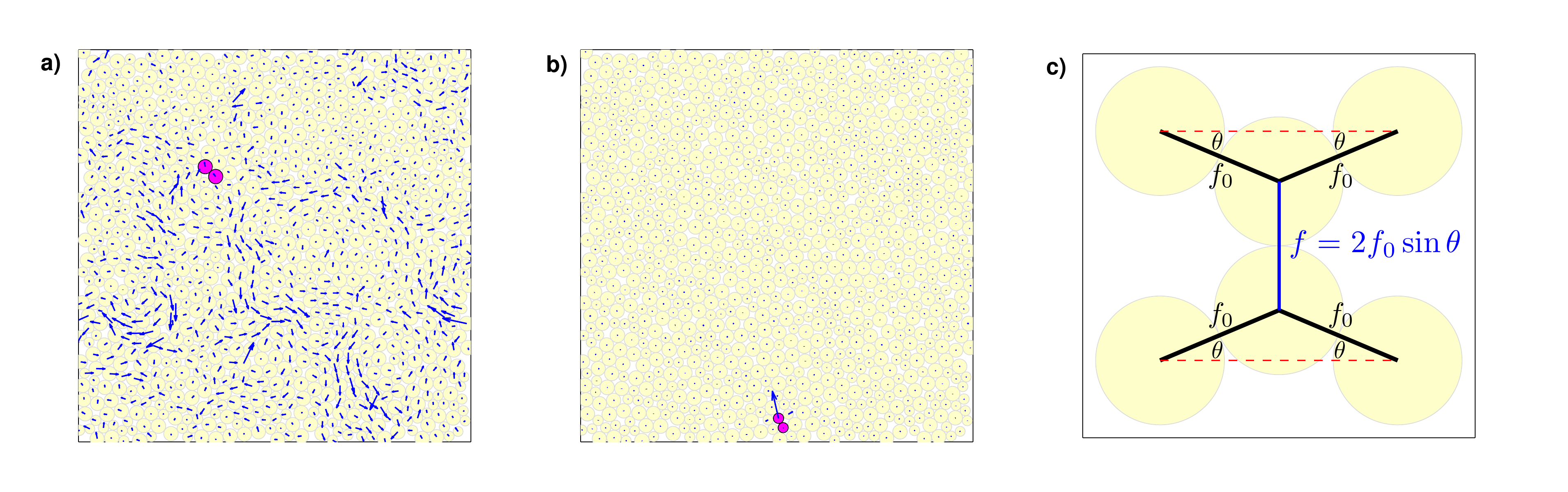}
\caption{
Examples of two  floppy modes, i.e.~displacement field resulting from pushing apart a pair of particles carrying a weak contact force, represented as
shaded. Panel {\bf a)} exemplified the case where  the displacements of the rest of the particles are of the same order of the
displacements of the pushed pair (corresponding in our notation to $b\sim1$), and the contact force is small because the floppy mode is almost orthogonal to a compression of the box.  Panel {\bf b)} displays a different displacement field generated in the same configuration as in panel {\bf a)}
by pushing apart a different pair of particles. This time the pushed pair's displacement is significantly larger than the displacements in the rest 
of the system, i.e.~$b\ll 1$. Such contacts are very weakly coupled to external stresses applied on the boundary, i.e.~they are mechanically isolated. Panel {\bf c)} displays a local configuration of particles that gives rise to small displacements when opening the vertical contact. Even if $f_0 \sim \langle f \rangle$, the force in the vertical contact can be small if the angle $\theta$ is small, and displacements resulting from opening that contact will be of order $b\sim \sin\theta$.}
\label{leverFig}
\end{figure*}

Floppy modes can be generated as follows \cite{tkachenko2}: two particles $1$ and $2$, forming a contact labelled $\alpha$, are pushed apart while all the other contacts remain closed, as represented in Figs.~\ref{pushedFig} and~\ref{leverFig}. We denote by $\delta {\vec R}^{(\alpha)}_{i}(s)$ the displacement of particle $i$ following the opening of the contact $\alpha$ by a distance $s$. This displacement field is uniquely defined, because only one floppy mode appears when a contact is broken, and exists for sufficiently small openings $s$, so as to ensure that no new contacts are formed in the system. It satisfies the following equation, that embodies the fact that contacts $\langle\generalPair\rangle$ other than $\alpha$ remain closed:
\begin{equation}
\label{foo05}
\delta\vec{R}^{\pair}_{\generalPair}(s)\cdot\vec{n}_{\generalPair} +{\cal O}(s^2)
= s\delta_{\alpha,\generalPair}\ ,
\end{equation}
where $\delta_{\alpha,\generalPair} = 1$ if and only if the \emph{pair} $\langle\generalPair\rangle$ is equal to the \emph{pair} $\alpha$, and is zero otherwise, and $\vec{n}_\generalPair \equiv \vec{R}_\generalPair/r_\generalPair$ is the director 
pointing from particle $i$ to particle $j$.

On the other hand, force balance in the unperturbed packing can be written as:
\begin{equation}
 \label{1}
 \forall i, \ \  {\vec F}_i+\sum_{ j(i)} f_{ij} {\vec n}_{ij}=0\ ,
\end{equation}
where the sum is on all particles $j(i)$ in contact with $i$, ${\vec F}_i$ is the force exerted by the wall on particle $i$ (and is thus zero for particles in the bulk), and $f_{ij}>0$ is the magnitude of the (purely repulsive) force in the contact $\langle ij \rangle$. Multiplying Eq.~(\ref{1}) by any displacement field $\delta {\vec R}_i$ and summing on all particles leads to the virtual work theorem:
\begin{equation}
\label{2}
\sum_i {\vec F}_i\cdot \delta {\vec R}_i+\sum_{\langle ij \rangle} \delta {\vec R}_{ij}\cdot{\vec n}_{ij} f_{ij}=0 \ .
\end{equation}
where $\sum_{\langle ij \rangle}$ denotes the summation over all contacts $\langle ij \rangle$. 
In our system external forces only stem from the boundaries, and the work associated with the displacement field 
$\delta {\vec R}$ is $\sum_i {\vec F}_i\cdot\delta {\vec R}_i=-p\delta \Omega$. 

Applying Eq.~(\ref{2}) with the vector field  $\delta {\vec R}^{(\alpha)}_{i}(s)$, we obtain \cite{theory}:
 \begin{equation}
 \label{foo01}
p\delta \Omega^{(\alpha)}\!=\!-\sum_i {\vec F}_i\cdot\delta {\vec R^{(\alpha)}}_i= sf_\alpha
+{\cal O}(s^2).
\end{equation}
Introducing the linear floppy mode
\begin{equation}\label{floppyMode}
{\vec V^{(\alpha)}_i}\equiv \left.\frac{d {\vec R^{(\alpha)}_i}}{d s}\right|_{s=0}\ ,
\end{equation}
one obtains by derivation:
\begin{equation}
\label{3first}
f_\alpha=-\sum_i {\vec F}_i\cdot{\vec V^{(\alpha)}_i}\ .
\end{equation}
It is convenient to introduce the vector field ${\vec V^*{}^{(\alpha)}_i}$ corresponding to the restriction of the floppy modes to the boundary, i.e.~${\vec V^*{}^{(\alpha)}_i}={\vec V^{(\alpha)}_i}$ for all particles $i$ on the boundary, and ${\vec V^*{}^{(\alpha)}_i}=0$ otherwise. Then we may write:
\begin{equation}
\label{3}
f_\alpha=-\sum_i {\vec F}_i\cdot{\vec V^*{}^{(\alpha)}_i}= -||F||\cdot ||V^{*(\alpha)}|| \cos(\theta_\alpha)
\end{equation}
where $||X||^2\equiv \sum_i {\vec X_i}^2$, and $\theta_\alpha$ is the angle made between the vector fields ${\vec F}_i$ and ${\vec V^*{}^{(\alpha)}_i}$. Eq.~(\ref{3}) indicates that the amplitude of a contact force is governed by how the  floppy mode associated to that contact couples to the external forces at the boundaries. 

A contact force can thus be small for two reasons. First, the amplitude of the displacement of the floppy mode are small, i.e.~$||V^{*\pair}||\ll \langle||V^{*(\beta)}||\rangle_\beta$ where $\langle \rangle_\beta$ denotes averaging over all contacts. For a typical contact it was shown that in an isostatic system ${\vec V^{(\alpha)}_i}\sim 1$ \cite{Wyart053}, as illustrated in Fig.~\ref{leverFig},a. However in some cases the local organization of the particles around the contact $\alpha$ is such that local displacements are weakly coupled to the rest of the system, as exemplified in Fig.~\ref{leverFig},b where significant motion occurs only near the perturbed contact. A local model of particle organization generating such a weak coupling with the rest of the system is sketched in Fig.~\ref{leverFig},c. 
To quantify this behavior we introduce the quantity:
\begin{equation}
\label{3bis}
b_{\alpha}\equiv ||V^{*(\alpha)}||/ \langle||V^{*(\beta)}||\rangle_\beta\ .
\end{equation}
$b_\alpha$ characterizes the floppy mode in the far field, which is of magnitude $||{\vec V^{(\alpha)}_i}||\sim b_\alpha$.
Note that contacts satisfying $b_\alpha \ll 1$ are mechanically isolated:  changes of external forces applied on the boundaries have little effect on the force amplitude, as implied by Eq.~(\ref{3}).  

The second possibility is that the floppy mode is nearly orthogonal to the external forces, i.e.~$\cos(\theta_\alpha)\ll \langle \cos(\theta_\beta)\rangle_\beta$. This situation is illustrated in Fig.~\ref{leverFig},a. To take these possibilities into account it is convenient to introduce $f_{typ}\equiv ||F||  \langle||V^{*(\beta)}||\rangle_\beta \langle \cos(\theta_\beta)\rangle_\beta$, and the dimensionless quantities:
\begin{equation}
\label{3ter}
 W_\alpha\equiv -\cos(\theta_\alpha)/\langle \cos(\theta_\beta)\rangle_\beta. 
\end{equation}
With these notations Eq.~(\ref{3}) can be rewritten as:
\begin{equation}
\label{4}
\frac{f_\alpha}{f_{typ}}=b_\alpha W_\alpha\ .
\end{equation}
As it has been shown \cite{Liu2} that the distribution of contact forces in packing decay at least exponentially fast at large forces, one thus expects that the distributions $P(b)$ and $P(W)$ rapidly decay when their arguments are larger than the typical values, i.e.~for $b>1$ and $W>1$. At small arguments we assume power-law forms:
\begin{eqnarray}
P(b)&\sim& b^\theta\ \ \hbox{      for }  b\ll 1\notag \\ 
P(W)&\sim& W^{\theta'} \hbox{ for }  W\ll 1\label{5}
\end{eqnarray}
In what follows we shall make the hypothesis that these two variables are independent $P(b,W)\approx P(b)P(W)$ (see numerical validation of this hypothesis in Appendix~\ref{independence}). It is then straightforward to compute the behavior of the force distribution at low-forces:
\begin{equation}
\label{6}
P(f)=\int db \, dW P(b) P(W) \delta (f-bW) \sim f^{\min(\theta,\theta')}.
\end{equation}
From Eq.~(\ref{6}) we learn that only the smaller of the two exponents
$\theta$ and $\theta'$ can be extracted from the distribution of contact forces $P(f)$. 

We have shown in this Section that weak contact forces $1 \gg f_\alpha \sim b_\alpha W_\alpha $ have two possible origins: either the displacements that result from opening a contact are small, i.e.~$b_\alpha\ll 1$, or the same displacements are weakly coupled to a compressive strain, i.e.~$W_\alpha \ll 1$ (or both). In the following we will show how pushing against contacts carrying weak forces may lead to rearrangements of the contact network, and how the response to this perturbation is related to the stability of jammed packings. We will argue that the exponents $\theta$ and $\theta'$ characterize two distinct pathways for structural rearrangements, and relate them to several aspects of the geometry and structure of packings. 

\section{Nonlinear stability analysis}
\label{stabilityCriterion}
\begin{figure}[ht]
\includegraphics[scale = 0.4]{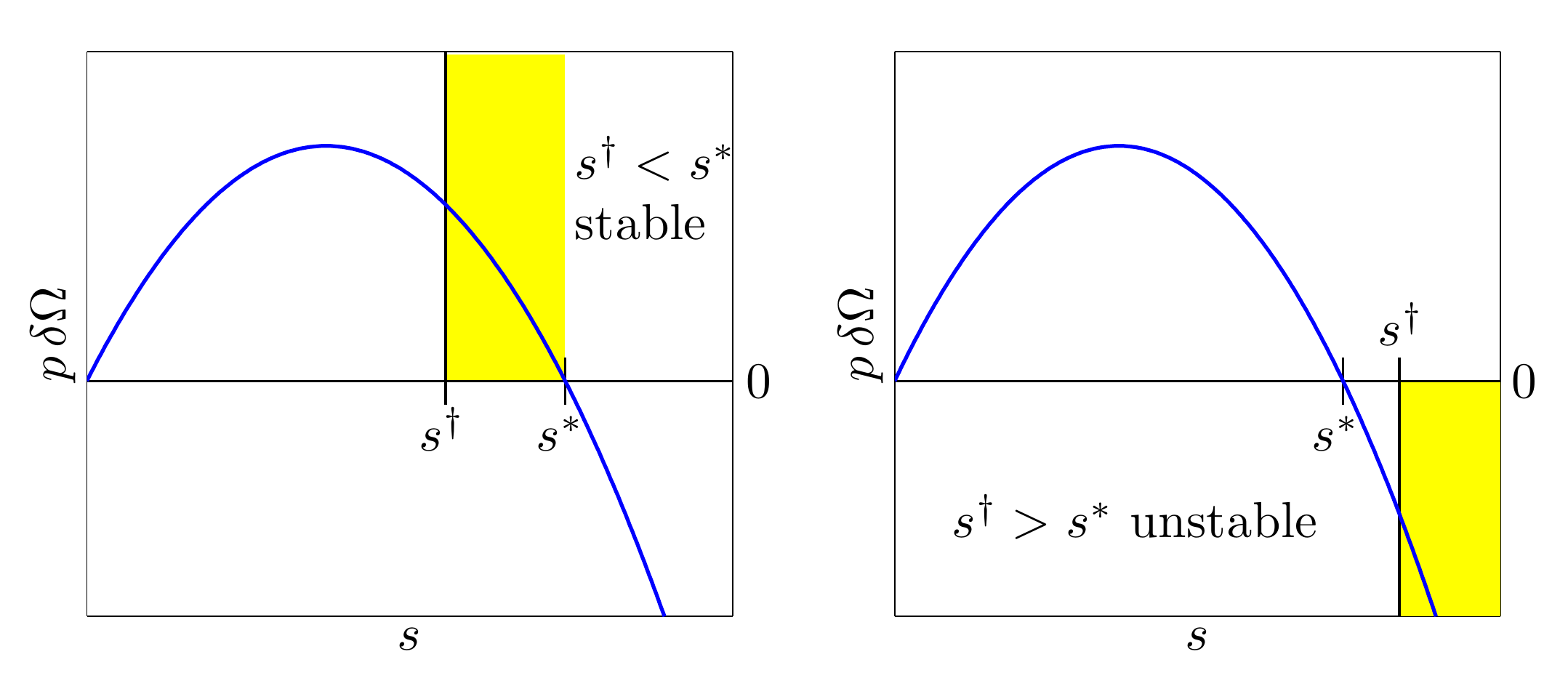}
\caption{Illustration of Eq.~(\ref{foo03}): the energy change $p\delta \Omega$ of opening a contact $\alpha$ by a distance $s$.
If the resulting displacement is blocked at an opening of $s^\dagger < s^*$ due to the creation of a new contact elsewhere in the system,
the packing is stable against opening the contact $\alpha$. On the other hand, 
if the contact $\alpha$ can be opened to distances $s > s^*$, a \emph{denser} packing can be found, and thus the packing 
would be unstable against pushing apart the contact $\alpha$.}
\label{illustration}
\end{figure}

In this Section we consider the stability of the system toward rearrangements of the contact network, in particular the possibility that the volume can be \emph{decreased} by opening a contact and pushing the system along a floppy mode (see Figs.~\ref{pushedFig} and~\ref{leverFig}). We consider again the displacements $\delta\vec{R}^\pair(s)$ that emanate from pushing apart the contact $\alpha$. Requiring that all other contacts are maintained in tact, and keeping terms up to second order in $s$, we arrive at the following equation for the displacements $\delta\vec{R}^\pair(s)$:
\begin{equation}\label{foo02}
\delta\vec{R}^{\pair}_{\generalPair}(s)\cdot\vec{n}_{\generalPair} +
\frac{\left(\delta\vec{R}^{\pair}_{\generalPair}(s)\cdot \vec{n}^\perp_{\generalPair}\right)^2}{2r_\generalPair}+{\cal O}(s^3)
= s\delta_{\alpha,\generalPair}\ ,
\end{equation}
where $\delta\vec{R}^{\pair}_{\generalPair}\cdot \vec{n}^\perp_{\generalPair}$ indicates the projection of
$\delta\vec{R}^\pair_\generalPair$ onto the space orthogonal to $\vec{n}_\generalPair$. Eq.~(\ref{foo02}) is analogous to Eq.~(\ref{foo05}), in addition to the inclusion of second order terms in $s$. We next apply Eq.~(\ref{2}) with the displacement field $\delta\vec{R}^\pair(s)$ defined by Eq.~(\ref{foo02}) and rearrange, to obtain:
\begin{equation}\label{foo06}
p\delta\Omega^\pair = sf_\alpha - 
\sum_{\langle ij \rangle} \sFrac{\left(\delta\vec{R}^\pair_\generalPair\cdot \vec{n}^\perp_\generalPair\right)^2\! f_\generalPair}{2r_\generalPair} + {\cal O}(s^3)\ .
\end{equation}
We introduce the dimensionless positive number:
\begin{equation}\label{cDefinition}
c_\alpha \equiv \lim_{s\rightarrow 0} \frac{\sum_{\langle ij\rangle}\! \frac{\left(\delta\vec{R}^\pair_\generalPair\cdot \vec{n}^\perp_\generalPair\right)^2f_\generalPair}
{2r_\generalPair}}{N\langle f \rangle s^2/a_0}=\frac{\sum_{\langle ij\rangle}\! \frac{\left(\vec{V}^\pair_\generalPair\cdot \vec{n}^\perp_\generalPair\right)^2f_\generalPair}
{2r_\generalPair}}{N\langle f \rangle /a_0} > 0\, ,
\end{equation}
and may then rewrite Eq.~(\ref{foo06}) as:
\begin{equation}\label{foo03}
p\delta \Omega = sf_\alpha - s^2c_\alpha N\langle f \rangle/a_0 + {\cal O}(s^3)\ .
\end{equation}
As the contact force $f_\alpha$ is positive, Eq.~(\ref{foo03}) implies that for a sufficiently small opening $s$ the volume always increases, as illustrated in Fig.~(\ref{illustration}). However, for larger openings the nonlinear term 
becomes important, and in particular if the opening $s$ satisfies:
\begin{equation}
s > s^* \equiv \frac{f_\alpha a_0}{c_\alpha\langle f\rangle N}\ ,
\end{equation}
a \emph{denser} packing can be generated \cite{theory}.
The distance $s$ by which a contact $\alpha$ can be opened is bounded however, since the displacements
of the particles in the entire system must eventually lead to the formation of a
new contact somewhere in the system. We denote by $s^\dagger$ the opening created between the pushed pair $\alpha$ when a new contact is formed. If $ s^\dagger< s^*$, stability is achieved: no denser packing can be generated along the direction of the floppy mode considered, as illustrated in Fig.~\ref{illustration}.

We define the \emph{stability index} $\kappa_\alpha$ as:
\begin{equation}
\label{foo04}
\kappa_\alpha \equiv \frac{f_\alpha a_0}{c_\alpha s^\dagger N\langle f \rangle}\ .
\end{equation}
If $\kappa_\alpha<1$, the packing is unstable against pushing against the contact $\alpha$, as $p\delta \Omega^\pair < 0$ according to Eq.~(\ref{foo03}),
implying that a denser packing can be generated. Eq.~(\ref{foo04}) can be used numerically to test for the stability of any contact in a packing, as shown below.

\section{Scaling relations and marginal stability}
\label{marginalStability}

The stability of the system requires that $\kappa_\alpha>1$ for all contacts $\alpha$. To unravel the consequence of this constraint, 
we perform a scaling analysis of Eq.~(\ref{foo04}). Firstly, we shall use that $f_\alpha\sim W_\alpha b_\alpha$ according to Eq.~(\ref{4}). 

Secondly, the quantity $c_\alpha$, defined as a normalized sum in Eq.~(\ref{cDefinition}), has two contributions.  There is a set of ${\cal O}(1)$ particles very close to the opened contact that satisfy $||{\vec V_i^\pair}||\sim 1$. Assuming that these particles also participate in contacts carrying forces of order $\langle f \rangle$ (as exemplified in Fig.~\ref{leverFig},c), they should lead to a contribution of order $1/N$ to $c_\alpha$.   The contribution from the bulk is of order $b^2$, since far away from the opened contact $||{\vec V_i^\pair}||\sim b$ and the majority of contact forces are of order $\langle f \rangle$. We may thus write $c_\alpha\sim1/N + b^2$ (a notation implying that $c_\alpha\sim 1/N$ if $b\ll1/\sqrt N$ and $c_\alpha\sim b^2$ otherwise).

Thirdly, the maximal displacement $s^\dagger$ that the pair $\alpha$ can be pushed apart before the creation of a new contact occurs is governed by 
the \emph{minimal gap} $h_{\rm min}$ between particles which are not in contact. In particular, a contact must be created when the motion of the particles in the bulk, of typical amplitude $b_\alpha s^\dagger$, is of order of $h_{\rm min}$,  leading to  $b_\alpha s^\dagger \sim h_{\rm min}$. 
The minimal gap follows:
\begin{equation}
\int_0^{h_{\rm min}}g(h')dh' \sim 1/N\ ,
\end{equation}
where $g(h)$ is the distribution of gap between particles.  Assuming a power-law behavior of the distribution of gaps, i.e.~$g(h) \sim h^{-\gamma}/a_0^{1-\gamma}$, one obtains $s^\dagger \sim h_{\rm min}/b_\alpha \sim (a_0N^{-\frac{1}{1-\gamma}})/b_\alpha$.

\begin{figure}[ht]
\includegraphics[scale = 0.5]{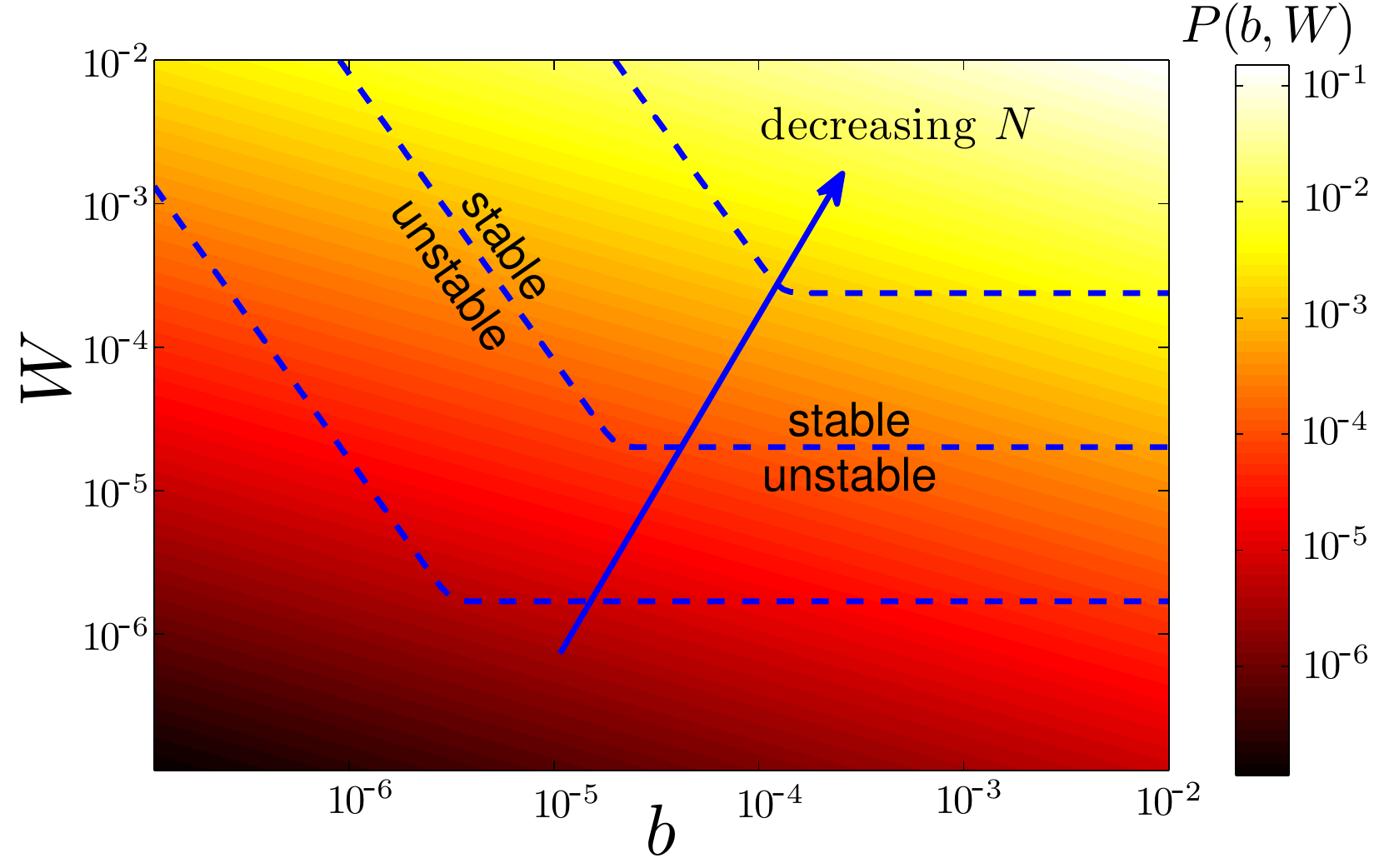}
\caption{Stability of contacts against opening: the color map corresponds to contours having equal values of 
$P(b,W)\approx P(W)P(b)$, which are assumed to take power-law forms, see Eq.~(\ref{5}). The dashed lines are deduced from Eqs.~(\ref{7}) and~(\ref{7a}), and correspond to the boundary separating stable contacts $\alpha$ having $\kappa_\alpha \ge 1$, to unstable contacts having $\kappa_\alpha \le 1$, for different system sizes $N$.}
\label{stabilityPhaseDiagram}
\end{figure}

Putting these results together, Eq.~(\ref{foo04}) becomes:
\begin{equation}
\label{7}
\kappa_\alpha\sim \frac{Wb^2 N^\frac{\gamma}{1-\gamma}}{\frac{1}{N}+b^2}\ .
\end{equation}
It is useful to examine the stability of contacts in the $(b$-$W)$ space; following Eq.~(\ref{7}), contacts $\alpha$ satisfying:
\begin{equation}\label{7a}
W_\alpha > \frac{b_\alpha^2 + N^{-1}}{b_\alpha^2N^\frac{\gamma}{1-\gamma}}\ ,
\end{equation}
are stable against opening. In Fig.~\ref{stabilityPhaseDiagram} a contour plot of the joint probability 
$P(b,W)$ is displayed, together with the stability limit given by Eq.~(\ref{7a}), for various system sizes $N$. 

Imposing the stability of all contacts requires that the probability of having an unstable contact in the system is much smaller than one, or equivalently:
\begin{equation}
\label{8}
\int_0^1P(\kappa)d\kappa\leq\frac{1}{N}\Leftrightarrow \!\!\!\!\!\!\!\int\limits_{\frac{Wb^2 N^\frac{\gamma}{1-\gamma}}{\frac{1}{N}+b^2}<1}\!\!\!\!\!\!\! dW db P(W) P(b) \leq\frac{1}{N}
\end{equation}
Decomposing the integration domain in two parts $b\ll1/\sqrt N$ and $b\gg1/\sqrt N$ one finds that the integral has two positive contributions, and that inequality~(\ref{8}) together with Eq.~(\ref{5}) is satisfied iff:
\begin{eqnarray}
\gamma&\geq&\frac{1-\theta}{2} \ , \label{9bis} \\
\gamma&\geq&\frac{1}{2+\theta'}\ . \label{9}
\end{eqnarray}
Each of the bounds (\ref{9bis}) and ~(\ref{9}) can be seen as resulting from the stability of a distinct relaxation process.

{\bf Mode 1} corresponds to contacts whose floppy modes are extended in the system, i.e.~$b\sim 1$ as illustrated in Fig.~\ref{leverFig}.a. 
The minimal contact force amplitude $f_{\min,1}$ encountered for such contacts in a system of size $N$ is simply $f_{\min,1}\sim W_{\min} \sim N^{-1/(\theta'+1)}$.
The reduction of volume stemming from non-linearities is dominated by the bulk, where the displacements are of order of the smallest gap $h_{\min}$ in the system, and $c_\alpha\sim 1$. Thus following Eq.(\ref{7}) the stability index of the weakest contact of that type is $\kappa\sim N^{\gamma/(1-\gamma)-1/(1+\theta')}$, and stability requires $\gamma\geq\frac{1}{2+\theta'}$. These modes strongly couple to changes in the forces at the boundary, and are expected to dominate the plastic  response occurring, for example, when a shear stress is applied, see Section \ref{discussion}.
%

{\bf Mode 2} corresponds to contacts whose associated floppy modes have small displacements in the far field, i.e.~$b\ll1$ as illustrated in Fig.~\ref{leverFig}.b,c. 
Mode~2 contacts turn out to be more numerous that mode 1 contacts at low forces, and thus dominate $P(f)$ at small $f$.  Their minimal force is thus the minimal force in the packing and follows $f_{\min,2}\sim b_{\min} \sim N^{-1/(\theta+1)}$. For these contacts the non-linear effect leading to a reduction of volume is dominated by the local displacements near  contact, i.e.~$c\sim 1/N$, whereas in the bulk the displacements are much smaller and of order $h_{\min}$. Thus mode 2  corresponds to a local buckling event of a few particles. This local buckling is, however, stabilized by the creation of a new contact in the far field. Eq.~(\ref{7}) for the weakest contact of that sort leads to $\kappa\sim N^{1+\gamma/(1-\gamma)-2/(1+\theta)}$, and stability requires $\gamma\geq\frac{1-\theta}{2}$. Mode 2 contacts are weakly coupled to external forces, and the frequency at which they yield under an increasing applied shear stress is much smaller that Mode 1, see Discussion. 

In the following we analyze numerically the structure and geometry of jammed packings of hard spheres. We test the relations between the exponents $\gamma, \theta, \theta'$ given by inequalities~(\ref{9bis}) and~(\ref{9}), and find that they are approximately saturated. This supports that the distribution of contact forces and the pair distribution function are coupled at random close packing. In addition, we verify the predicted existence of two distinct modes of relaxation, namely mode 1 and mode 2, and demonstrate their marginal stability.

\section{Results from computer experiments}
\label{testingRelation}
In this Section we numerically test the predictions made in Section~\ref{marginalStability}, namely the relations between the scaling exponents $\gamma, \theta$ and $\theta'$ given in Eqs.~(\ref{9bis}) and~(\ref{9}), and assess the stability of jammed packings. We produced an ensemble of jammed packings of various system sizes, and directly measured all of the structural and geometrical quantities discussed in the previous sections. For a detailed description of the numerical methods and calculations used in this work, see Appendix~\ref{appendix}. 

\begin{figure}[!ht]
\includegraphics[scale = 0.5]{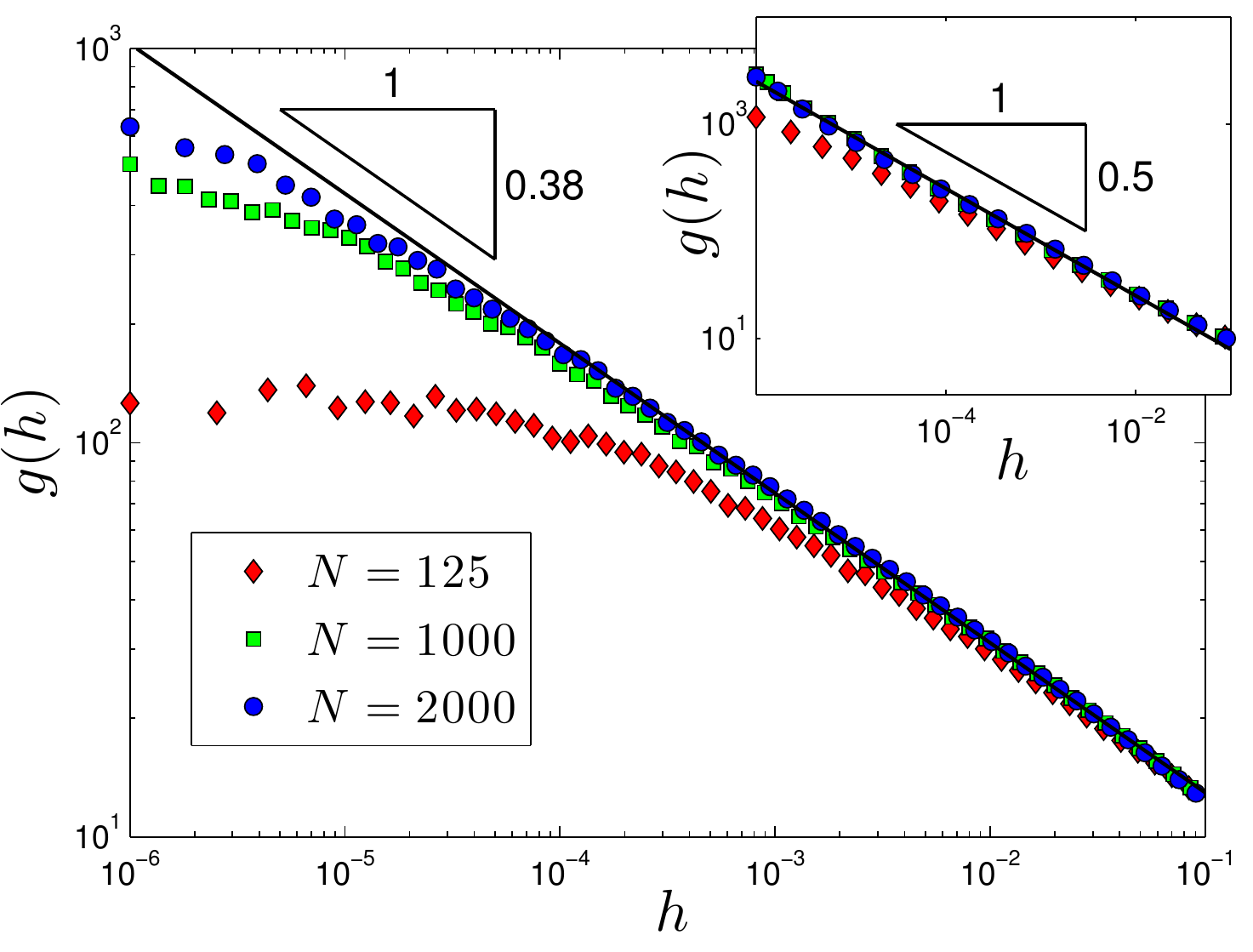}
\caption{Distributions of gaps between particles which are not in contact, measured in our ensemble of jammed packings for different system sizes. The continuous line represent the power law $g(h) \sim h^{-0.38}$. \underline{Inset:} The same distributions of gaps measured in jammed configurations \emph{before} eliminating rattlers. The tails of the distributions are different compared to the those of the rattler-free jammed packings, and seem to follow $g(h) \sim h^{-0.5}$ (which is represented by the
continuous line in the inset). This means that rattler-free packings have much less small gaps compared to packings with rattlers, as previously known \cite{Silbert06,donev2}.}
\label{gapsFigure}
\end{figure}

\subsection{Scaling relations}

We begin our analyses with the distribution of gaps between particles which are not in contact, $g(h)$, which is plotted in Fig.~\ref{gapsFigure}. 
This quantity is computed after removing the small fraction of rattlers that do not belong to the rigid structure. This is the relevant observable in our problem, as explained in Appendix B.
We observe strong finite size effects in the shape of the distributions in the small gaps range: the power law behavior seems to break down
at some system-size dependent gap which is decreasing with increasing the system size. 
In the range in which the distributions agree, they obey a power law. The divergence of the distribution at small gaps follows $g(h) \sim h^{-\gamma}$ with $\gamma \approx 0.38$.
We note, however, that larger jammed configurations are required to improve our estimation of this exponent. 

\begin{figure}[ht]
\includegraphics[scale = 0.52]{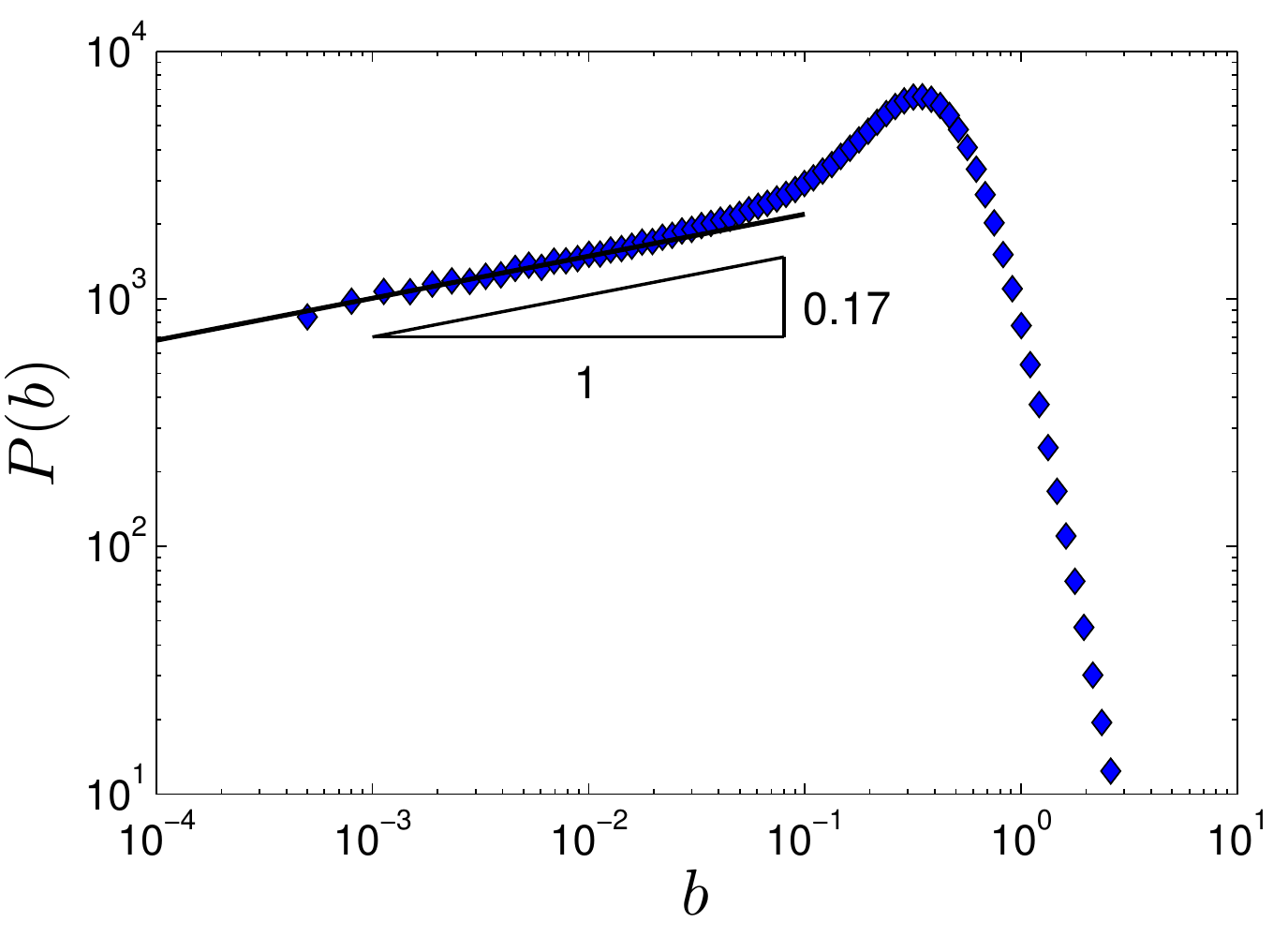}
\caption{Distribution $P(b)$ characterizing the mechanical decoupling of contacts. }
\label{P(b)}
\end{figure}

We next turn to the distribution of $b_\alpha$ (see Eq.~(\ref{3bis}) for definition) which represents the magnitude of the far-field displacements of the floppy mode emanating from pushing apart the contact $\alpha$, see Appendix~\ref{appendix} for details about the numerical calculations. Fig.~\ref{P(b)} presents the distribution $P(b)$ measured in our jammed packings of size $N=1000$. We indeed find that $P(b) \sim b^\theta$ follows a power law for small $b$, with $\theta \approx 0.17$.

\begin{figure}[ht]
\includegraphics[scale = 0.52]{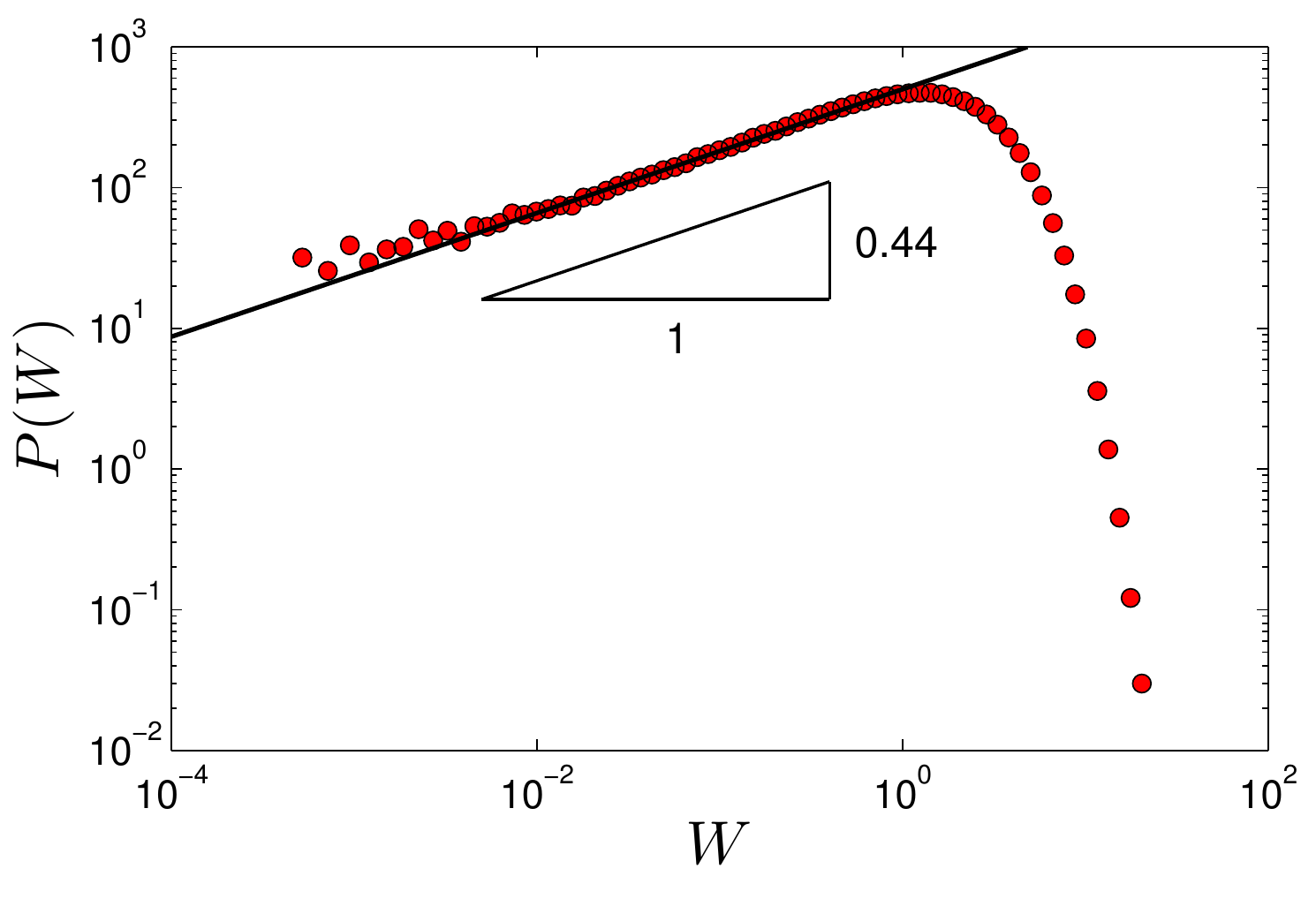}
\caption{Distribution $P(W)$ characterizing the angle made between a floppy mode and a compressive strain. }
\label{P(W)}
\end{figure}

In Figure~\ref{P(W)} we plot the distribution of the couplings $W_\alpha$ of the floppy modes $\vec{V}^\alpha$ to a compressive strain,
defined in Eq.~(\ref{3ter}). Here too we find that $P(W) \sim W^{\theta'}$ follows a power law for small $W$, with $\theta' \approx 0.44$. 

\begin{figure}[!ht]
\includegraphics[scale = 0.55]{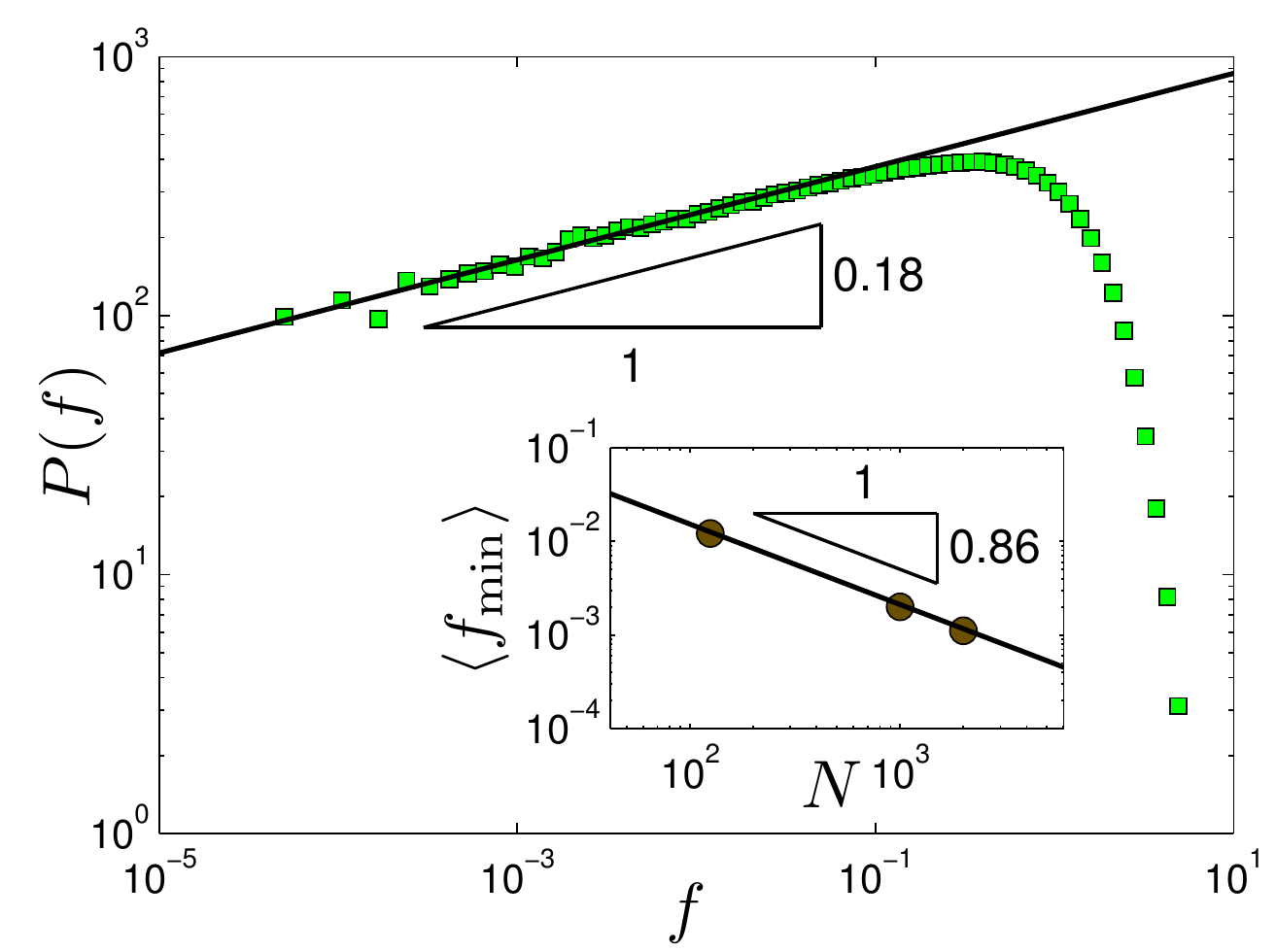}
\caption{Distribution $P(f)$ of contact forces $f$ measured in our ensemble of jammed packings of systems with $N=1000$ particles,
under a pressure $p = 1$. We do not observe any systematic system-size dependence in the same distributions measured for different $N$. Inset: the mean minimal contact force in a packing vs.~packing size $N$.}
\label{contactForceDistribution}
\end{figure}

\begin{figure*}[!ht]
\includegraphics[scale = 0.47]{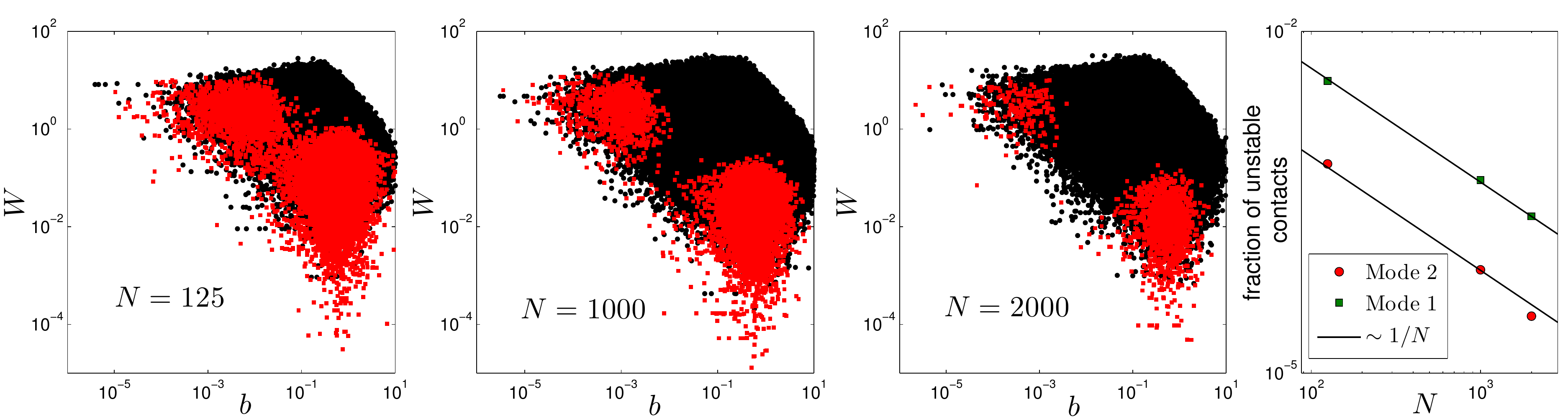}
\caption{Stability of contacts plotted in the $(b$-$W)$ plane - red squares represent contacts that, when perturbed, lead to a denser packing, whereas black circles represent stable contacts. The regions of instability and their system-size dependence is consistent with our scaling analysis as displayed in Fig.~\ref{stabilityPhaseDiagram}, and clearly indicate the existence of two distinct types of instabilities. We note that although mode 2 contacts are more numerous than mode 1 contacts (and therefore dominate the distribution of contact forces at low force), the number of \emph{unstable} mode 2 contacts is \underline{smaller} than the number of \emph{unstable} mode 1 contacts, as manifested in the different prefactors of the scaling of the fraction of unstable contacts, plotted in the rightmost panel.}
\label{redPlot}
\end{figure*}

With numerical measurements for the scaling exponents $\gamma,\theta$ and $\theta'$ in hand, we now turn to validate relations~(\ref{9bis}) and~(\ref{9}); for Eq.~(\ref{9bis}) we find $0.38\approx \gamma\geq(1-\theta)/2\approx 0.41$, whereas for Eq.~(\ref{9}) we find
$0.38 \approx \gamma \ge 1/(2+\theta') \approx 0.41$. The very slight violations of the bounds are within the numerical errors, and our measures support that the bounds are in fact saturated. 

In Fig.~\ref{contactForceDistribution} we plot the distribution of the contact forces $P(f)$ in our jammed packings. We find
(as shown before in \cite{edan2} for jammed configurations obtained during shear flows) that the distribution behaves at small forces as $P(f) \sim f^\theta/\langle f\rangle^{\theta+1}$ with $\theta\approx 0.18$, a power-law that appears to hold nicely on three decades. 
Another test of this exponent appears in the inset of Fig.~\ref{contactForceDistribution} that displays the dependence of the mean minimal force with system size $N$.
We observe $\langle f_{\min}\rangle\sim N^{-0.86}$, corresponding to $\theta= 1/0.86-1 \approx 0.16$, in good agreement with our direct fit. These findings validate our prediction spelled out in Eq.~(\ref{6}) according to which $P(f) \sim f^{\min(\theta,\theta')}$. In Appendix~\ref{independence} we demonstrate the independence of the variables~$b$ and~$W$, which is assumed to make the prediction of Eq.~(\ref{6}).

Note that $P(f)$ and $g(h)$ depend on the protocol by which  jammed packings are generated, but apparently not on the spatial dimension~\cite{12CCPZ,10CBS}. For hard sphere packings obtained via a Stillinger algorithm Charbonneau et al.~\cite{12CCPZ} found that $P(f)\sim f^{0.28}$ and $g(h) \sim h^{-0.42}$, i.e. $\gamma \approx 0.42$, and to $\theta=0.28$ assuming that $\theta<\theta'$. This allows us to test Eq.~(\ref{9bis}): $0.42\approx \gamma\geq (1-\theta)/2\approx 0.36$; i.e.~a satisfied bound but again quite close to saturation. 

For soft decompressed particles these authors found $P(f)\sim f^{0.42}$ and $ \gamma\approx 0.39$. If $\theta<\theta'$ the bound of Eq.~(\ref{9bis}) gives $0.39\approx \gamma\geq (1-\theta)/2\approx 0.29$, i.e.~a satisfied bound but now somewhat further away from marginality. Note however that since the exponent they find for the force distribution is large  in this case, and close to our value of $\theta'$ (see below), we suspect that the condition $\theta<\theta'$ does not hold in this case. If $\theta'<\theta$ then the bound Eq.~(\ref{9}) should be used instead, and  reads $0.39\approx \gamma\geq 1/(2+\theta')\approx 0.41$, i.e.~consistent with the marginality of mode one, as we also found with very similar exponents.

\subsection{Distinct modes of instability}
With the measurements of $b_\alpha, s^\dagger_\alpha, c_\alpha$, and~$W_\alpha$ in hand, we can directly measure the stability index $\kappa_\alpha$ of each contact $\alpha$ in each jammed packing of our ensemble via Eq.~(\ref{foo04}). The stability of contacts on the $(b$-$W)$ plane is visualized in Fig.~\ref{redPlot}. We find a consistent picture to that described in Fig.~\ref{stabilityPhaseDiagram}. Two distinct modes of relaxation emerge from this analysis: mode 1, which are contacts $\alpha$ having $b_\alpha\sim1$ and $W_\alpha\ll1$, and mode 2, which are contacts $\alpha$ having $b_\alpha\ll1$ and $W_\alpha\sim1$. This result demonstrates that contacts carrying small forces $f_\alpha\sim b_\alpha W_\alpha$ are potentially unstable, though the mechanical nature of the instability can belong to either one of the two distinct modes. 

We found that roughly two contacts in a packing, if opened, may lead to a denser packing, and are therefore unstable. Accordingly, we find that the fraction of unstable contacts, plotted in the rightmost panel of Fig.~(\ref{redPlot}), scales as $1/N$ for Mode~I and~II respectively, indicating that our packings are indeed marginally stable. This further vindicates our findings that the inequalities presented in Eqs.~(\ref{9bis}) and~(\ref{9}) appear to be saturated. 


\section{Avalanches of plastic events}
\label{discussion}

When Eqs.~(\ref{9bis}) and~(\ref{9}) are strictly satisfied, for large packings one  never observes instabilities: in that case $s^\dagger \ll s^*$ even for the contacts carrying the weakest forces, implying according to Fig.~\ref{illustration} that opening a gap between particles simply leads to a linear increase of volume. However, empirically we observe that the bounds of Eqs.~(\ref{9bis}) and~(\ref{9}) appear saturated, and as shown in Fig.~\ref{redPlot} we find that there are instabilities in our packings, both for type~1 and type~2 contacts. Thus if temperature is present, or if shear stress is applied, such instabilities will occur, and one must consider the rate at which these events take place, and their consequences.  

We shall consider the case where external forces (corresponding, e.g.,~to a shear stress) are applied. According to Eq.~(\ref{3first}) applying an external force ${\vec \Delta F_i}$ on some particle $i$  changes the contact force in some contact $\alpha$ by an amount \cite{tkachenko2}:
\begin{equation}
\label{11}
\Delta f_\alpha= {\vec V^{(\alpha)}_i}\cdot {\vec \Delta F_i}\sim b_\alpha \Delta F_i\ .
\end{equation}
For a contact to yield its force must vanish, i.e.~$\Delta f_\alpha\leq -f_\alpha=b_\alpha W_\alpha f_{typ}$, thus the threshold stress where a rearrangement occurs is $\Delta F\sim W_\alpha f_{typ}$: it is large for type~2 contacts for which $W_\alpha\sim 1$, but very small for type~1 contacts where $W_\alpha$ can be vanishingly small. This argument supports that mode~1 rearrangements are much more frequent and presumably govern the plastic response under an imposed stress. 
 
Let us assume that mode~1 and/or mode~2 are marginally stable, and consider the consequence of their respective relaxation. 
Let $\beta$ label the contact that closes at $s^\dagger$. By symmetry,  the results obtained when the contact $\alpha$ was opened and $\beta$ was closed also apply to the newly obtained configuration if $\beta$ is re-opened and $\alpha$ is re-closed. In particular, the relation
\begin{equation}
f_\alpha= \left.\frac{\partial }{\partial s_\alpha}\left[p\delta V(s_\alpha)\right]\right|_{s_\alpha=0}
\end{equation}
becomes
\begin{equation}
f_\beta= \left.\frac{\partial }{\partial s_\beta}\left[p\delta V(s_\beta)\right]\right|_{s_\beta=0}
\end{equation}
where $s_\beta$ is the distance by which the contact $\beta$ is opened. Since both for modes 1 and 2, the contact $\beta$ is far from the contact $\alpha$, one has  $\partial s_\beta/\partial s_\alpha\sim b_\alpha$. Furthermore,  marginal stability corresponds to $s^\dagger\sim s^*$ for the weakest contacts. This implies that 
\begin{equation}
\left.\frac{\partial }{\partial s_\alpha}\right|_{s_\alpha=0}\left[p\delta V(s_\alpha)\right] \sim
\left.\frac{\partial }{\partial s_\alpha}\right|_{s_\alpha=s^\dagger}\left[p\delta V(s_\alpha)\right] \ ,
\end{equation}
from which we get:
\begin{equation}
\label{12}
f_{\beta}= \left.\frac{\partial }{\partial s_\alpha}\left[p\delta V(s_\alpha)\right]\right|_{s_\alpha=s^\dagger}
\frac{\partial s_\alpha}{\partial s_\beta}\sim \frac{f_\alpha}{b_\alpha}
\end{equation}
Thus the force in the new contact is of order of the contact that was opened for mode 1 relaxation for which $b\sim{\cal O}(1)$, but it is much larger for mode 2, for which $b$ can be vanishingly small.
Forming a new contact is equivalent to applying a  dipole of external force of the system.
According to Eq.~(\ref{11}), this will change the contact force in any contact $\delta$ by some amount $\Delta f_\delta\sim f_\alpha (b_\delta/b_\alpha)$. 

Consider that a type 1 instability occurs, i.e.~$f_\alpha\sim f_{1,\min}$ and $b_\alpha\sim 1$. The effect on the weakest type 2 contacts, labeled $\delta =(2,{\min})$ is negligible as in that case
\begin{equation}
\Delta f_{2,\min}\sim f_{1,\min} b_\delta \sim \frac{f_{1,\min}}{f_{typ}} f_\delta \ll f_{2,\min}\ .
\end{equation}
However the weakest type 1 contacts has a finite probability to become unstable and to open, as $\Delta f_{1,\min}\sim f_{1,\min}$. Thus when the system is marginally stable, the relaxation of a Mode 1 contact can lead to a sequence of  events where other such contacts become unstable and open. Numerically such avalanches of relaxation are seen and are power-law distributed \cite{rouxDL}, a situation that is only possible if marginal stability is satisfied \cite{theory}.

The same argument applied when a type 2 contact relaxes leads to a similar conclusion: the marginal stability of Mode 2 corresponds precisely to the situation where the relaxation of a type 2 contact can trigger the instability of another such contact with a finite probability. However one finds a key difference: if a mode 2 contact relaxes it will affect significantly all the  contacts forces in the system, thus resulting in a large global change of the contact network. Thus Mode 2 relaxation appears to be extremely rare and hard to trigger, but to have a very dramatic effect when it occurs. 

\section{Soft particles}
\label{softParticles}
Packings of hard particles are isostatic, and the response to a local strain does not decay with distance, unlike generic amorphous materials in which elastic responses to local strains decay as power laws away from the perturbation.
It is thus important to consider how our results extend to more realistic models, for example to soft compressed particles ($\phi>\phi_c$) or thermal colloidal systems ($\phi<\phi_c$). 
Do the relaxations processes that we have found in packings of hard particles still exist in such systems, and if so what fixes their density?

Previous works \cite{brito07b,brito3,manning2,xuepl} aiming at relating relaxation processes to the microscopic structure in these systems have focused on the linear response, in particular on the
 soft vibrational  modes very different from plane waves that are present at low-frequency. The presence of such modes is expected if these systems live close to an elastic instability, as is indeed observed after numerical quenches into the glass phase are performed, at least for a rather large interval of packing fractions around $\phi_c$ \cite{O'Hern03,Wyart052,brito07a}.  The idea that nearly unstable modes would generate low-activation barriers above which the system can relax is rather natural, and is supported by numerical observations showing that thermal relaxation mostly occurs along a few soft modes both in colloidal systems \cite{brito07b} and in Lennard-Jones systems \cite{hocky}. An accurate description  of the barriers associated with soft modes is however lacking \cite{footnote1}  and would be very interesting to investigate numerically.

The present work suggests that when the interaction potential is non-linear (for example  soft spheres where the potential is a one-sided harmonic spring), intrinsically non-linear relaxation processes should be considered too. 
 This view that the relaxation processes of type~1 and/or type~2 play a role away from the jamming threshold is supported by the recent numerical results of Charbonneau et al.~\cite{12CCPZ}, where $P(f)$ and $g(h)$ have been carefully measured as the packing fraction is varied, both in colloidal systems $(\phi<\phi_c)$ and in systems of soft particles $(\phi > \phi_c)$. For concreteness, we shall focus on soft  particles, which in that study are harmonic, and where the stiffness of the interactions is taken to be unity.  These authors find that under a compression of amplitude $\Delta \phi$, power-law tails are still observed in the distribution functions of structural quantities, which in our notation writes
\begin{equation}
P(f)\sim \left(\frac{f}{\langle f\rangle}\right)^{\!\!\min (\theta,\theta')}\!\!\!\!\!\!\sim 
\left(\frac{f}{\Delta\phi}\right)^{\!\!\min (\theta,\theta')}\!\!\!\!, \ \mbox{and}\ g(h)\sim \left(\frac{h}{a_0}\right)^{\!\!-\gamma}\!\!\!,
\end{equation}
excepted on an {\it extremely narrow} interval of gaps $h<h^*\sim \Delta \phi^\mu$ and forces $f<f^*\sim \Delta \phi^\mu$ with $\mu\approx 1.75$, where these distributions do not significantly vary. As pointed out in~\cite{12CCPZ}, the exponent $\mu$ can be related to $\gamma$ and to the force exponent (${\min (\theta,\theta')}$ in our notation). To see this, let us generalize the  definition of the gap distribution function $g(h)$ to include for the possibility of negative gaps, i.e.~particles in contact for which $f_\alpha=-h_\alpha$. 
Assuming that $g(h)\sim 1/h^\gamma$ for $h \ll\Delta \phi^\mu$ and $g(h)\sim (-h)^{\min (\theta,\theta')}/ \Delta \phi^{\min (\theta,\theta')}$ (as required by the force distribution) for $-h \gg-\Delta \phi^\mu$, and that $g(h)$ smoothly varies for $|h|\ll\Delta \phi^\mu$ implies by continuity that $\mu=(1+{\min (\theta,\theta')})/(\gamma+{\min (\theta,\theta')})$. If $(\theta<\theta')$ marginal stability of mode two (see Eq.~(\ref{9bis})) implies $\mu=2$, whereas if $\theta'<\theta$ marginal stability of mode one would imply $\mu=(1/\gamma-1)/(\gamma+1/\gamma-2)\approx 1.64$,  in closer agreement with the observations of~\cite{12CCPZ}.
 
The observations that force and pair distributions are nearly unchanged in compressed packings, excepted on a very narrow range of forces and gaps, suggest that the modes we introduced still play a role in controlling the structure in this situation.  The numerical data of \cite{12CCPZ} are consistent with the idea that occurrences of mode 2 are rare in compressed packing, perhaps because they are too fragile and ``buckle" under compression.  To test this scenario and distinguish which of the exponents $\theta$ or $\theta'$ is smaller in soft spheres, one would need to perform a numerical analysis similar to the one presented in this work.

 \section{Conclusion}
\label{conclusion}

We conclude by a summary of our results and open questions. We have established the existence of two distinct non-linear modes of relaxation in random packings of hard frictionless spheres, associated with changes of the network of contacts. Mode two corresponds to the local buckling of a few particles.  Such excitations are numerous and their density is described by an exponent $\theta$  that also characterizes the distribution of contact forces $P(f)\sim f^\theta$. Using a stability argument we proved that the density of such modes is bounded from above, leading to $\gamma\geq\frac{1-\theta}{2}$, where $\gamma$ characterizes the singularity of the pair distribution function $g(r)$ at contact. Mode one, whose density is characterized by the exponent $\theta'$, corresponds to extended motion of the particles, and the bound on their density is $\gamma\geq\frac{1}{2+\theta'}$. We performed numerics that support that these bounds are saturated with $\gamma=0.38$, $\theta=0.17$ and $\theta'=0.44$. In addition, we introduced numerical methods to analyze the entire sets of excitations of a given packing, which indicate that the two modes of relaxation considered are indeed barely stable. These results support that marginality  is the fundamental principle governing which ensemble of configurations are visited in this out-of-equilibrium system, and constrain key aspects of the microscopic structure of packings, such as the distribution of force $P(f)$ at low forces and the pair distribution function $g(r)$ at small gaps. It also yields a natural explanation for the crackling dynamics observed in driven packings of hard frictionless particles. 
%
%
%
%

Concerning the random close packing of hard particles, three questions stand out. ($i$) If marginal stability is indeed found, as appears to be the case for our system preparation,  the three exponents $\theta$, $\theta'$ and $\gamma$ are related by two relationships.  What then, fixes the exponent $\gamma$? ($ii$)  What explains the difference between soft and hard spheres, and are mode two marginal or even present in soft spheres as well?
($iii$) Hard sphere packings display power-law distributed avalanches of plasticity~\cite{rouxDL}. Can the exponent characterizing the size of the power-law be computed? 
 
Next it is important, as discussed in Section~\ref{softParticles},  to extend this approach to situations where  particles are elastic and compressed or to colloidal glasses.
The structure still appears to be sensitive to the relaxations modes we have introduced. Is there still some kind of marginality present? What fixes the concentration of 
these relaxation processes?

Lastly, there is evidence that these relaxation processes play an important role in flow. Anisotropic configurations that jammed during flow display a distribution of force nearly identical to isotropic packings \cite{edan2}.
Away from the jamming threshold this distribution is cut off below some relative force scale that vanishes near jamming \cite{edan2}, similarly to what happens in soft particles or colloidal systems \cite{12CCPZ}. These results suggest to develop a description of contact dynamics and its effect on the rheological properties of dense suspensions based on these excitations \cite{inProgress}. 

\begin{acknowledgments}
We thank  A.~Grosberg and P.~Hohenberg for discussions, and Noa Lahav for help with the graphics. This work has been supported primarily by the MRSEC Program of the National Science Foundation
DMR-0820341, by the Sloan Fellowship,
by the National Science Foundation DMR-1105387, and by the Petroleum Research Fund
\#52031-DNI9.
\end{acknowledgments}

\appendix
\section{Independence of $b$ and $W$}
\label{independence}
The quantities $b$ and $W$, defined in Eqs.~(\ref{3bis}) and~(\ref{3ter}) respectively, are assumed to be statistically independent in the range $b\ll1$ and $W\ll1$. This assumption is used when the prediction of the force distribution form is made in Eq.~(\ref{6}), and in the scaling arguments made for Eq.~(\ref{8}). In this Appendix we demonstrate how well this assumption holds by analyzing data from our numerical simulations. For a detailed description of the numerical methods and of calculations of structural quantities, see Appendix~\ref{appendix}. 
\begin{figure}[!ht]
\includegraphics[scale = 0.55]{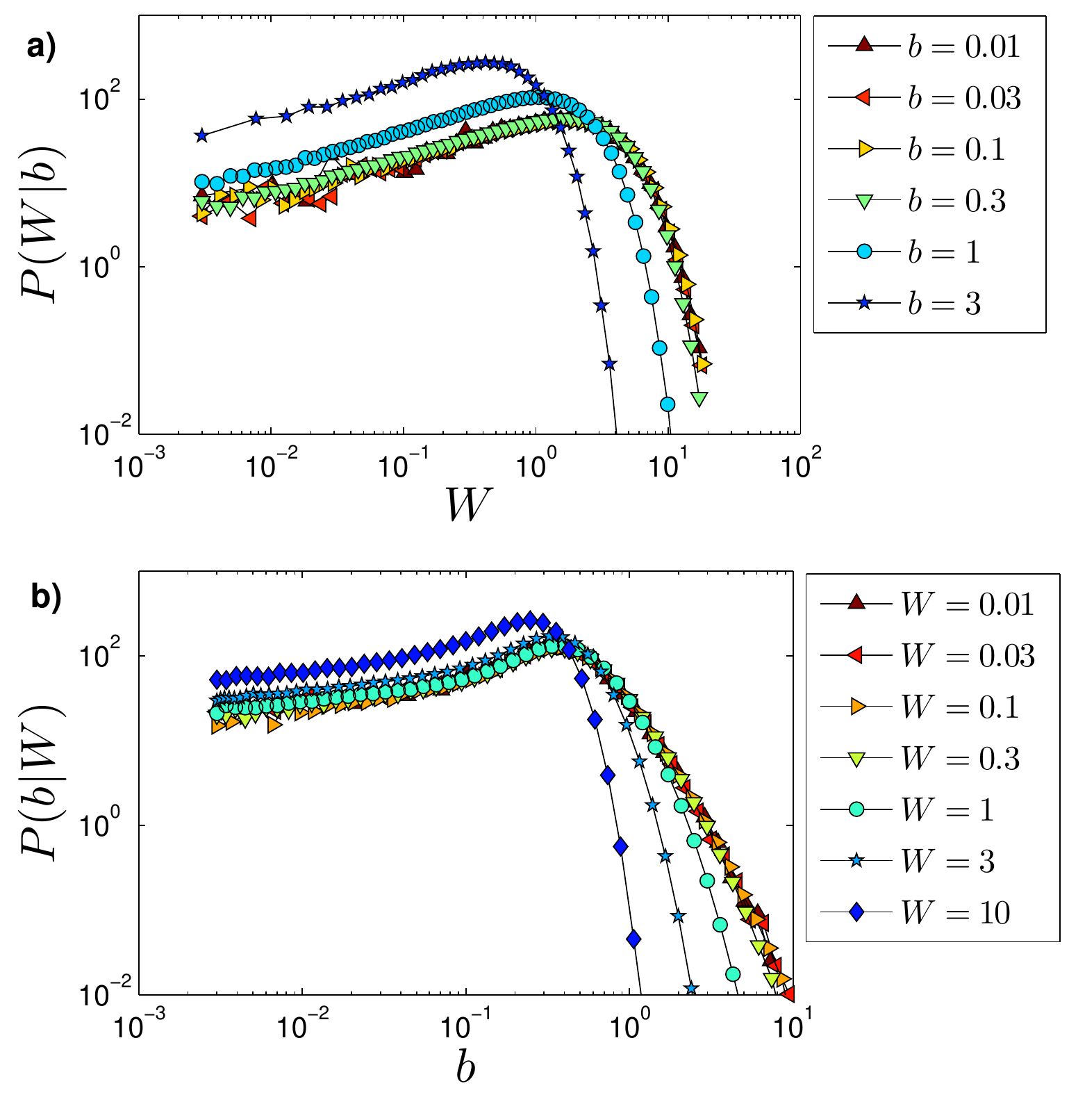}
\caption{Conditional distributions: {\bf a)} $P(W|b)$, {\bf b)} $P(b|W)$. Both conditional distributions seem to converge to a limiting distribution when the conditioned variable is smaller than unity.}
\label{independenceFig}
\end{figure}

If $b$ and $W$ were independent, one would observe that the conditional probabilities $P(b|W)$ and~$P(W|b)$ obey
\begin{equation}
P(b|W) = P(b) \ \ \mbox{and}\ \ P(W|b) = P(W)\ ,
\end{equation}
namely that the conditional distribution does not depend on the value of the conditioned variables. In Fig.~\ref{independenceFig} we exemplify that for $b<<1$ and $W<<1$ this is indeed the case. We conclude from these results that in these limits, the joint distribution $P(b,W) \approx P(b)P(W) \sim b^\theta W^{\theta'}$, see Eq.~(\ref{5}). 

\section{Numerical methods}
\label{appendix}
To generate numerically jammed packings under periodic boundary condition we used the event-driven method we developed in \cite{numericPaper}; the method enables one to isotropically compress systems of hard frictionless spheres under overdamped dynamics.  We initialized each independent compression run with a random initial condition
at a packing fraction $\phi_{\rm initial} < \phi_c \approx 0.645$ lower than the critical packing fraction $\phi_c$, 
and compressed the systems until they jammed. We generated about 1000 jammed configurations
for $N=2000$, about 4500 jammed configurations for $N=1000$ and about 9000 jammed configurations for $N=125$. 
In the compression simulations we bound the mean error in contacts overlap to be smaller than
$10^{-8}a_0$~\cite{numericPaper}, where $a_0$ is the mean diameter of the particles. We employed a 50:50 binary mixture of large and small particles,
such that the ratio between the radii of the large and small particles is~1.4.

In jammed configurations, a small percentage of the particles are loosely connected to the rigid backbone of the packing~\cite{numericPaper}.
We eliminated these `rattler' particles by fixing the position of the all the particles
belonging to the rigid backbone, and applying a `gravitational' force field such that the rattlers sink to the bottom of their respective 
cavities, and create at least three contacts with neighboring particles. We generalized the same method described in~\cite{numericPaper} for this step of our numerics. Eliminating the rattlers  turns out to be important to study quantitatively the  stability of packings -- although how they are removed, by sinking or by taking them out of the system, leads to the same results.  Packings in which rattlers are not eliminated posses many more small gaps compared to rattler-free packings.
These small gaps would affect our numerical estimation of the stabilization distances $s^\dagger$ where new contacts are formed. Thus, the amplitude of the displacements 
$\delta\vec{R}^{\pair}$ resulting from pushing apart contacts would appear more limited than what they actually are, since new contacts created with rattlers do not confer mechanical stability. 

The structural quantities of interest for the analysis described in this work are the contact forces $f_\alpha$ defined in Eq.~(\ref{1}), the magnitude of the far-field displacements $b_\alpha$ defined in Eq.~(\ref{3bis}), the cosine of the angles $W_\alpha$ as defined in Eq.~(\ref{3ter}), and the stabilization distances $s^\dagger_\alpha$ at which a new contact is formed. The extraction of these quantities from our jammed packings is explained in the following bulleted paragraphs:

\begin{itemize}
\item{Contact forces are resolved by solving Eq.~(\ref{1}), together with the constraint that the pressure is fixed:
\begin{equation}
\label{pressure}
\sum_{\langle ij \rangle} f_\generalPair r_\generalPair  = p\,\Omega\, d\ .  
\end{equation}
}
\item{Calculation of the far-field displacements $b_\alpha$ requires the availability of the floppy mode $\vec{V}^\pair$ which emanates from pushing against a contact~$\alpha$, defined in Eq.~(\ref{floppyMode}). A jammed packing embedded in a fixed container possesses no such modes, and in the arguments of Sect.~\ref{marginalStability} we let the walls confining the system move, thus allowing for the volume to vary. The same situation occurs with periodic boundary conditions, where the system size needs to be allowed to change length to generate one floppy mode by opening one single contact. We thus let the periodic system change length, or metric, and define $e^{\pair}(s) = (L(s)-L(0))/L(0)$, the compressive strain that occurs as the contact $\alpha$ is opened by $s$. 
In this geometry the floppy mode equation is not Eq.~(\ref{foo02}), but rather: 
\begin{equation}
\label{softMode}
\delta\vec{R}^{\pair}_{\generalPair}\cdot\vec{n}_{\generalPair} + 
\frac{\left(\!\delta\vec{R}^{\pair}_{\generalPair}\cdot \vec{n}^\perp_{\generalPair}\!\right)^2}{2r_\generalPair}
+ e^\pair(s) r_{\generalPair} = s\delta_{\alpha,\generalPair} + {\cal O}(s^3).
\end{equation}
From Eq.~(\ref{softMode}) it is straightforward to derive  our previous results that remain unchanged, in particular Eqs.~(\ref{foo06}) and~(\ref{foo04}). 
Numerically, we solve the equation
\begin{equation}
\vec{V}^{\pair}_{\generalPair}\cdot\vec{n}_{\generalPair} + \frac{de^\pair(s)}{ds} r_{\generalPair} = \delta_{\alpha,\generalPair}
\end{equation}
for the floppy mode $\vec{V}^\pair$ associated with the opening of any contact $\alpha$. With the floppy mode $\vec{V}^\pair$ in hand, we can directly calculate the dimensionless constants $c_\alpha$ as defined in Eq.~(\ref{cDefinition}).

\begin{figure}[ht]
\includegraphics[scale = 0.44]{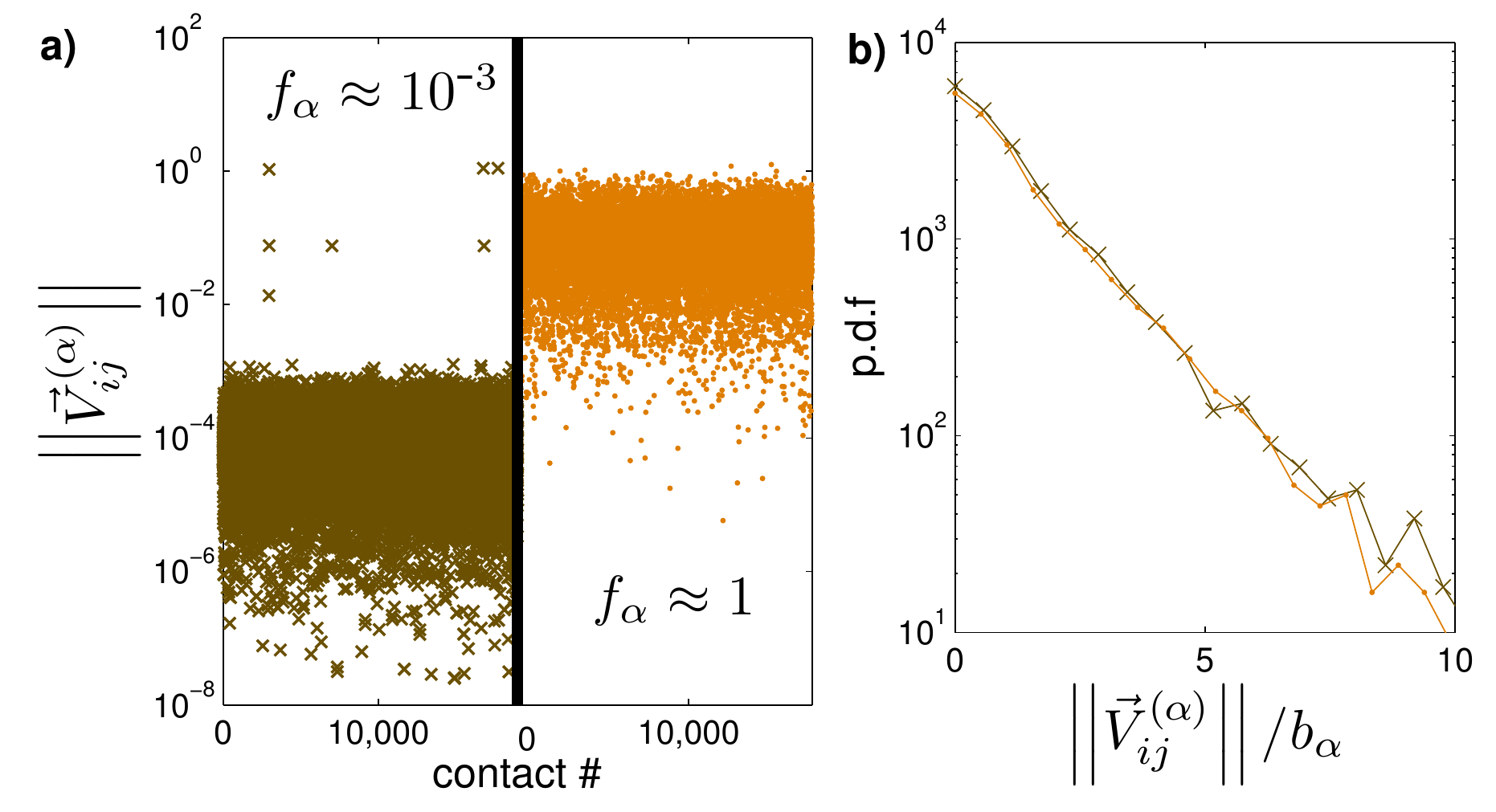}
\caption{{\bf a)} Scatter plot of relative displacements of particles in contact, following the pushing apart of a contact 
carrying a weak force $f_\alpha\approx 10^{-3}$ (crosses) and a typical force $f_\alpha\approx 1$ (dots), both measured in a single jammed configuration in two dimensions with $N=10000$ particles. Notice that even in the weak force case (crosses), there are a few contacts with relative displacement of order unity, which is why considering the median of relative displacements (and not the mean) is useful.
{\bf b)} Distributions of relative displacements rescaled by the medians $b_\alpha$ defined in Eq.~(\ref{leverDefinition}). The symbols
are consistent with the left panel.}
\label{demonstrate}
\end{figure}

The definition of $b_\alpha$ proposed in Eq.~(\ref{3bis}) is not practical in a periodic system,
and numerically we choose the following definition:
\begin{equation}\label{leverDefinition}
b_\alpha = \underset{\mbox{\tiny contacts $\langle ij \rangle$}}{\mbox{median}}\left\{ \left|\left| 
\vec{V}^\pair_\generalPair\right|\right|\right\}\ ,
\end{equation}
where the median is taken over all contacts. It is important to consider the median and not the mean of the relative displacements as the mean is affected by the few particles near contact for which ${\vec V}\sim 1$, and thus does not represent the amplitude of displacements in the far field. Fig.~\ref{demonstrate} shows  the distribution of relative displacements following the opening of a contact, and exemplifies its dependence on the strength of the contact force. These observations demonstrate that when  a contact $\alpha$ is opened and displacement along a floppy mode occurs, a well-defined displacement scale $b_\alpha$ emerges.
}
\item{
Once the far-field displacements $b_\alpha$ are calculated, then following Eq.~(\ref{4}) the cosine of the angles $W_\alpha$ of Eq.~(\ref{3ter}) are calculated as
\begin{equation}
W_\alpha = f_\alpha/b_\alpha\ .
\end{equation}
}
\item{
The stabilization displacements $s^\dagger_\alpha$ are the maximal distance $s$ that the pair $\alpha$ can be displaced before creating a new contact. To calculate $s^\dagger_\alpha$, we consider all pairs of particles $\langle\generalPair\rangle$ that are not in contact, and use the scheme for finding the next collision time in a hard sphere gas~\cite{91AT}, treating $\vec{V}^\pair$ as velocities and $s$ as time.
}

\end{itemize}


\begin{thebibliography}{99}

\bibitem{theory}
M.~Wyart, Phys.~Rev.~Lett.~{\bf 109}, 125502 (2012).

\bibitem{phillips_book}
W.~A.~Phillips, \emph{Amorphous Solids: Low-Temperature Properties}, Vol.~24 (Springer, Berlin, 1981).

\bibitem{langer98}
M.~L.~Falk and J.~S.~Langer, Phys.~Rev.~E {\bf 57}, 7192 (1998).

\bibitem{argon}
A.~Argon, Acta Metall.~{\bf 27}, 47 (1979).

\bibitem{04ML}
C.~Maloney and A.~Lema{\^i}tre, Phys.~Rev.~Lett.~{\bf 93}, 195501 (2004).

\bibitem{harrowell2}
A.~Widmer-Cooper, H.~Perry, P.~Harrowell, and D.~R.~Reichman, J.~Chem.~Phys.~{\bf 131}, 194508 (2009).

\bibitem{brito07b}
C.~Brito and M.~Wyart, J.~Stat.~Mech.~(2007) L08003.

\bibitem{manning2}
M.~L.~Manning and A.~J.~Liu, Phys.~Rev.~Lett.~{\bf 107}, 108302 (2011).

\bibitem{O'Hern03}
C.~S.~O'Hern, L.~E.~Silbert, A.~J.~Liu, and S.~R.~Nagel, Phys.~Rev.~E {\bf 68}, 101136 (2003).

\bibitem{donev2}
A.~Donev, S.~Torquato, and F.~H.~Stillinger, Phys.~Rev.~E {\bf 71}, 011105 (2005).

\bibitem{Silbert06}
L.~E.~Silbert, A.~J.~Liu, and S.~R.~Nagel, Phys.~Rev.~E {\bf 73}, 041304 (2006).

\bibitem{12CCPZ}
P.~Charbonneau, E.~I.~Corwin, G.~Parisi, F.~Zamponi, Phys.~Rev.~Lett.~{\bf 109}, 205501 (2012).

\bibitem{edan2}
E.~Lerner, G.~D{\"u}ring, and M.~Wyart, Europhys.~Lett.~{\bf 99}, 58003 (2012).

\bibitem{Liu2}
C.~H.~Liu, S.~R.~Nagel, D.~A.~Schecter, S.~N.~Coppersmith, S.~Majumdar, O.~Narayan, and T.~A.~Witten, Science 269, 513 (1995).

\bibitem{sno}
J.~H.~Snoeijer, T.~J.~H.~Vlugt, M.~van Hecke, and W.~van Saarloos, Phys.~Rev.~Lett.~{\bf 92}, 054302 (2004).

\bibitem{edwards}
S.~F.~Edwards and R.~B.~S.~Oakeshott, Physica A {\bf 157}, 1080 (1989).

\bibitem{torquato}
S.~Torquato and F.~H.~Stillinger, Rev.~Mod.~Phys.~{\bf 82}, 2633 (2010).

\bibitem{Zamponi}
G.~Parisi and F.~Zamponi, J.~Chem.~Phys.~{\bf 123}, 144501 (2005).

\bibitem{efros}
A.~L.~Efros and B.~I.~Shklovskii, J.~Phys.~C {\bf 8}, L49 (1975).

\bibitem{monroe}
D.~Monroe, A.~C.~Gossard, J.~H.~English, B.~Golding, W.~H.~Haemmerle, and M.~A.~Kastner, Phys.~Rev.~Lett.~{\bf 59}, 1148 (1987).

\bibitem{Pazmandi}
F.~P{\'a}zm{\'a}ndi, G.~Zar{\'a}nd, and G.~T.~Zim{\'a}nyi, Phys.~Rev.~Lett.~{\bf 83}, 1034 (1999).

\bibitem{goethe}
M.~Goethe and M.~Palassini, Phys. Rev. Lett.~{\bf 103}, 045702 (2009).

\bibitem{markus}
M.~Müller and S.~Pankov, Phys.~Rev.~B {\bf 75}, 144201 (2007).

\bibitem{thouless}
D.~Thouless, P.~Anderson, and R.~Palmer, Philos.~Mag.~{\bf 35}, 593 (1977).

\bibitem{moore}
P.~R.~Eastham, R.~A.~Blythe, A.~J.~Bray, and M.~A.~Moore, Phys.~Rev.~B {\bf 74}, 020406 (2006).

\bibitem{horner}
H.~Horner, Eur.~Phys.~J.~B {\bf 60}, 413 (2007).

\bibitem{crack}
J.~Sethna, K.~Dahmen, and C.~Myers, Nature {\bf 410}, 242 (2001).

\bibitem{Wyart052}
M.~Wyart, L.~E.~Silbert, S.~R.~Nagel, and T.~A.~Witten, Phys.~Rev.~E {\bf 72}, 051306 (2005).

\bibitem{Wyart053}
M.~Wyart, Annales de Phys.~{\bf 30} (3), 1 (2005).

\bibitem{brito07a}
C.~Brito and M.~Wyart, Europhys.~Lett.~{\bf 99}, 149 (2006).

\bibitem{brito3}
C.~Brito and M.~Wyart, J.~Chem.~Phys.~{\bf 131}, 024504 (2009).

\bibitem{brujic2}
I.~Jorjadze, L.-L.~Pontani, and J.~Brujic, Phys.~Rev.~Lett.~{\bf 110}, 048302 (2013).

\bibitem{Wyart05}
M.~Wyart, S.~R.~Nagel, and T.~A.~Witten, Europhys.~Lett.~{\bf 72}, 486 (2005).

\bibitem{wyart2010}
M.~Wyart, Europhys.~Lett.~{\bf 89}, 64001 (2010).

\bibitem{13DLW}
G.~D{\"u}ring, E.~Lerner, and M.~Wyart, Soft Matter {\bf 9}, 146 (2013).

\bibitem{vitelli2010}
V.~Vitelli, N.~Xu, M.~Wyart, A.~J.~Liu, and S.~R.~Nagel, Phys.~Rev.~E {\bf 81}, 021301 (2010).

\bibitem{Silbert05}
L.~E.~Silbert, A.~J.~Liu, and S.~R.~Nagel, Phys.~Rev.~Lett.~{\bf 95}, 098301 (2005).

\bibitem{respprl}
W.~G.~Ellenbroek, E.~Somfai, M.~van Hecke, and W.~van Saarloos, Phys.~Rev.~Lett.~{\bf 97}, 258001 (2006).

\bibitem{berthier13}
A.~Ikeda, L.~Berthier, and G.~Biroli, J.~Chem.~Phys.~{\bf 138}, 12A507 (2013).

\bibitem{revue}
W.~van Saarloos, M.~Wyart, A.~J.~Liu, and S.~R.~Nagel, \emph{The jamming scenario an introduction and outlook} (Oxford University Press, Oxford, UK), (2010). 

\bibitem{rouxDL}
G.~Combe and J.-N.~Roux, Phys.~Rev.~Lett.~{\bf 85}, 3628 (2000).

\bibitem{Schreck}
C.~F.~Schreck, T.~Bertrand, C.~S.~O'Hern, and M.~D.~Shattuck, Phys.~Rev.~Lett.~{\bf 107}, 078301 (2011).

\bibitem{alexander}
S.~Alexander, Phys.~Rep.~{\bf 296}, 65 (1998).

\bibitem{tkachenko2}
A.~V.~Tkachenko and T.~A.~Witten, Phys.~Rev.~E {\bf 62}, 2510 (2000).

\bibitem{moukarzel}
C.~F.~Moukarzel, Phys.~Rev.~Lett.~{\bf 81}, 1634 (1998).

\bibitem{10CBS}
P.~Chaudhuri, L.~Berthier, and S.~Sastry, Phys.~Rev.~Lett.~{\bf 104}, 165701 (2010).

\bibitem{xuepl}
N.~Xu, V.~Vitelli, A.~J.~Liu, and S.~R.~Nagel, Europhys.~Lett.~{\bf 90} (2010).

\bibitem{hocky}
G.~M.~Hocky and D.~R.~Reichman, arXiv:1211.0033 (2012).

\bibitem{footnote1}
Imposing a displacement along soft modes as performed in \cite{xuepl} is presumably not appropriate as the barrier must not be straight, imposing a force, perhaps in conjunction with the string method,  should already lead to a significant improvement.

\bibitem{inProgress}  G.~D{\"u}ring, E.~Lerner, and M.~Wyart, in progress.

\bibitem{numericPaper}
E.~Lerner, G.~D{\"u}ring, and M.~Wyart, Comp.~Phys.~Comm.~{\bf 184}, 628 (2013). 


\bibitem{91AT}
M.~P.~Allen and D.~J.~Tildesley, \emph{Computer Simulations of Liquids}
(Oxford Univ.~Press, New York, 1991). 






\end{thebibliography}
\end{document}